\documentclass[aps, prl, superscriptaddress,showkeys, preprint]{revtex4-1}
\usepackage{titlesec}
\usepackage{graphicx}
\usepackage{dcolumn}
\usepackage{bm}
\usepackage{color}
\usepackage{array}
\usepackage{tabularx}
\usepackage{tabularray}
\usepackage{multirow, makecell, booktabs}
\usepackage{threeparttable}

\usepackage{amsmath}
\usepackage{amssymb}
\usepackage{mhchem}
\usepackage{comment}
\usepackage{ragged2e} 

\usepackage{pbox}
\usepackage{float}
\DeclareUnicodeCharacter{0394}{$\Delta$}
\usepackage[colorlinks=true, linkcolor=blue, citecolor=blue, urlcolor=blue]{hyperref}

\usepackage{stmaryrd}  

\begin{document}


\preprint{}

\title{\textbf{Multimode magnon-phonon cavity driven by symmetry-locked strain fields}}

\author{Chunli Tang}
\altaffiliation{These two authors contributed equally}
\affiliation{Department of Physics, Auburn University, Auburn, AL 36849, USA}
\affiliation{Department of Electrical and Computer Engineering, Auburn University, Auburn, AL 36849, USA}

\author{Yujie Zhu}
\altaffiliation{These two authors contributed equally}
\affiliation{Department of Materials Science and Engineering, University of Wisconsin-Madison, Madison, WI 53706, USA}

\author{Dayne Sasaki}
\affiliation{Department of Materials Science and Engineering, University of California, Davis, Davis, CA 95616, USA}

\author{Jiaxuan Wu}
\affiliation{Department of Materials Science and Engineering, University of Wisconsin-Madison, Madison, WI 53706, USA}

\author{Yuzan Xiong}
\affiliation{Department of Physics and Astronomy, University of North Carolina at Chapel Hill, Chapel Hill, NC 27599, USA}

\author{Harshil Goyal}
\affiliation{Department of Physics, Auburn University, Auburn, AL 36849, USA}
\affiliation{Department of Electrical and Computer Engineering, Auburn University, Auburn, AL 36849, USA}

\author{Masoud Mahjouri-Samani}
\affiliation{Department of Electrical and Computer Engineering, Auburn University, Auburn, AL 36849, USA}

\author{Mark Adams}
\affiliation{Department of Electrical and Computer Engineering, Auburn University, Auburn, AL 36849, USA}

\author{Xiang Meng}
\affiliation{Department of Electrical Engineering, Columbia University, New York, NY 10027, USA}

\author{Bethany E. Matthews}
\affiliation{Environmental Molecular Sciences Laboratory, Pacific Northwest National Laboratory, Richland, Washington 99354, USA}

\author{Le Wang}
\affiliation{Environmental Molecular Sciences Laboratory, Pacific Northwest National Laboratory, Richland, Washington 99354, USA}

\author{Yingge Du}
\affiliation{Environmental Molecular Sciences Laboratory, Pacific Northwest National Laboratory, Richland, Washington 99354, USA}

\author{Jia-Mian Hu}
\affiliation{Department of Materials Science and Engineering, University of Wisconsin-Madison, Madison, WI 53706, USA}

\author{Yayoi Takamura}
\affiliation{Department of Materials Science and Engineering, University of California, Davis, Davis, CA 95616, USA}

\author{Wei Zhang}
\affiliation{Department of Physics and Astronomy, University of North Carolina at Chapel Hill, Chapel Hill, NC 27599, USA}

\author{Wencan Jin}
\email{wjin@auburn.edu}
\affiliation{Department of Physics, Auburn University, Auburn, AL 36849, USA}
\affiliation{Department of Electrical and Computer Engineering, Auburn University, Auburn, AL 36849, USA}

\date{\today}

\begin{abstract}

\noindent Hybrid magnon-phonon cavities with precise control knobs are highly sought after for coherent energy and signal transduction in solid-state platforms. While strain offers a powerful means to tune magnonic characteristics, extending strain engineering into magnon-phonon hybridization has remained elusive. Moreover, implementing controllable strain at the meso- or nanoscale poses a formidable challenge, as spatial inhomogeneity of strain fields often leads to enhanced damping and reduced coherence, thereby hindering device integration and scalability. Here, we present an epitaxial \ce{La_{0.7}Sr_{0.3}MnO3}/\ce{SrTiO3} (LSMO/STO) heterostructure that exhibits strong coupling between the Kittel magnon and acoustic phonon. Leveraging the emergence of structural domains when STO undergoes a cubic-to-tetragonal phase transition, we create anisotropic local strains at the interface. Remarkably, the anisotropic local strain of less than 0.1\% drives the pronounced splitting of the magnon into three branches. Each branch independently hybridizes with acoustic phonons, forming a matrix of magnon-phonon avoided crossings that underpins multimode transduction and programmable networks in frequency and magnetic field space. An analytical model reveals that the split magnon branches are deterministically locked to the three main crystalline axes, enabling robust, orientation-selective control of the hybridized magnon-phonon spectrum against spatial inhomogeneity. Our results establish designed local strain as an exceptionally sensitive trigger for multimode magnon-phonon hybridization in magnetoelastic oxide heterostructures, and highlight local strain engineering as a viable strategy for designing tunable hybrid magnonic and phononic devices.

\end{abstract}

\maketitle

Strong coupling between two quantized excitations in a cavity gives rise to hybridized states that bestow novel quantum phenomena. In particular, hybrid magnon–phonon cavities, in which collective spin and lattice excitations couple coherently, provide a powerful platform for exploring rich thermodynamic and spin transport phenomena in solids ~\cite{kikkawa2016magnon,man2017direct,ramos2019room,yahiro2020magnon}, as well as for developing quantum technologies based on coherent energy and information transduction ~\cite{li2020hybrid,bozhko2020magnon,li2021advances}. Strongly coupled magnon–phonon systems enable the formation of magnon polarons that manifest avoided crossings in the spectra, making them attractive for cavity magnonics and emerging quantum acoustic architectures. An overarching goal in this frontier is to realize multimode magnon–phonon hybridization, which is essential for implementing complex coupling networks, mode‑selective control, and scalable quantum functionalities ~\cite{o2010quantum,manenti2017circuit,moores2018cavity,PhysRevApplied.17.034024}.

To date, several strategies have been exploited to generate multimode magnon polarons, predominantly in magnetic multilayer systems. These include exciting higher‑order standing spin‑wave modes ~\cite{christy2025tuning,qin2018exchange}, exploiting nonlinear magnon processes ~\cite{xu2023magnonic,xiong2024magnon,shen2025cavity}, or integrating multiple magnetic layers within a single cavity ~\cite{PhysRevB.101.060407, PhysRevX.12.011060}. Despite these advances, a central challenge remains: is there a deterministic and controllable route to create multiple non-degenerate magnon modes and enable their independent hybridization with phonons, without relying on higher-order modes or increasing structural complexity?

Strain offers a powerful means to excite magnon dynamics and tune magnonic characteristics ~\cite{weiler2011elastically,weiler2012spin,scherbakov2010coherent,kim2012ultrafast,Zhuang2024HybridMagnonPhonon}. Although strain can be introduced through a variety of approaches such as lattice mismatch, patterned nanostructures, and surface acoustic waves, it is generically treated as a global and uniform field. Consequently, it primarily leads to rigid shifts of magnon dispersion or resonance frequency, while preserving magnon mode degeneracy. In contrast, local strain, which varies at meso- or nanoscales, remains far less explored, largely due to the experimental challenges of generating and characterizing it in a controlled manner. Nevertheless, local strain fields create a distinct landscape within a single material, leading to profound modification of optoelectronic and transport properties ~\cite{castellanos2013local,kalisky2013locally}, or stabilizing magnetic domains in composite multiferroics~\cite{yang2014elastically,hu2016multiferroic}. Moreover, it has been shown that local strain acts as an effective gauge field that corresponds to a magnetic field ranging from millitesla to 300 Tesla ~\cite{guinea2010energy,zhou2023imaging,karki2026quantum}. Motivated by these findings, we envisage that the designed local strain, which does not average out in the presence of spatial inhomogeneity, can serve as a symmetry-breaking field. Such local strain thus provides a unique yet unexplored route to lift magnon degeneracy and realize multimode magnon–phonon hybridization beyond the limitations of global strain control.

High-quality oxide heterostructures have enabled a wide range of magnetoelastic functionalities that can be engineered through their extreme sensitivity to interface effects \cite{zubko2011interface, hwang2012emergent}. In this work, we present an \ce{La_{0.7}Sr_{0.3}MnO3}/\ce{SrTiO3} (LSMO/STO) heterostructure that manifests a unique testbed for local strain-driven multimode magnon-phonon hybridization. LSMO is a half-metallic ferromagnetic oxide with large magnetoelastic coupling strength ~\cite{PhysRevB.84.184412} and low magnon damping rate ~\cite{PhysRevB.109.014437, liu2019current, zhang2021long}. It can be epitaxially grown on an STO substrate, a quantum paraelectric perovskite that undergoes a cubic-to-tetragonal phase transition at a critical temperature of $T_\mathrm{S}\sim$ 105 K ~\cite{STO_Structural}. In an LSMO/STO heterostructure with optimized magnon and phonon profiles, our ferromagnetic resonance (FMR) spectra show that efficient excitation and coherent coupling between the Kittel magnon mode and transverse acoustic (TA) phonons can be prompted near room temperature. Notably, across the cubic-to-tetragonal phase transition, the formation of three types of structural domains generates anisotropic local strains at the interface; as a consequence, the Kittel magnon splits into three branches, and each branch couples to the TA phonons separately with distinct coupling characteristics. An analytical model reveals that, despite spatial inhomogeneity, the local strains are rigorously locked to the [100], [010], and [001] crystalline orientations, thus modifying the magnetocrystalline anisotropy and lifting the degeneracy of the Kittel magnon. This mechanism exhibits exceptional sensitivity, functioning robustly even for anisotropic local strains below 0.1\%. While local strain is often regarded as a disorder whose effects average out macroscopically, we demonstrate that it can instead act collectively as an emergent symmetry-breaking field and exert a decisive influence on structure-property relationships. Moreover, our work highlights magnetoelastic oxide interfaces as a superior platform for creating and controlling local strain-driven multimode magnon-phonon hybridization, which is compatible with existing cryogenic quantum hardware for coherent transduction applications.

\begin{figure}[ht!]
\includegraphics[width=0.75\textwidth]{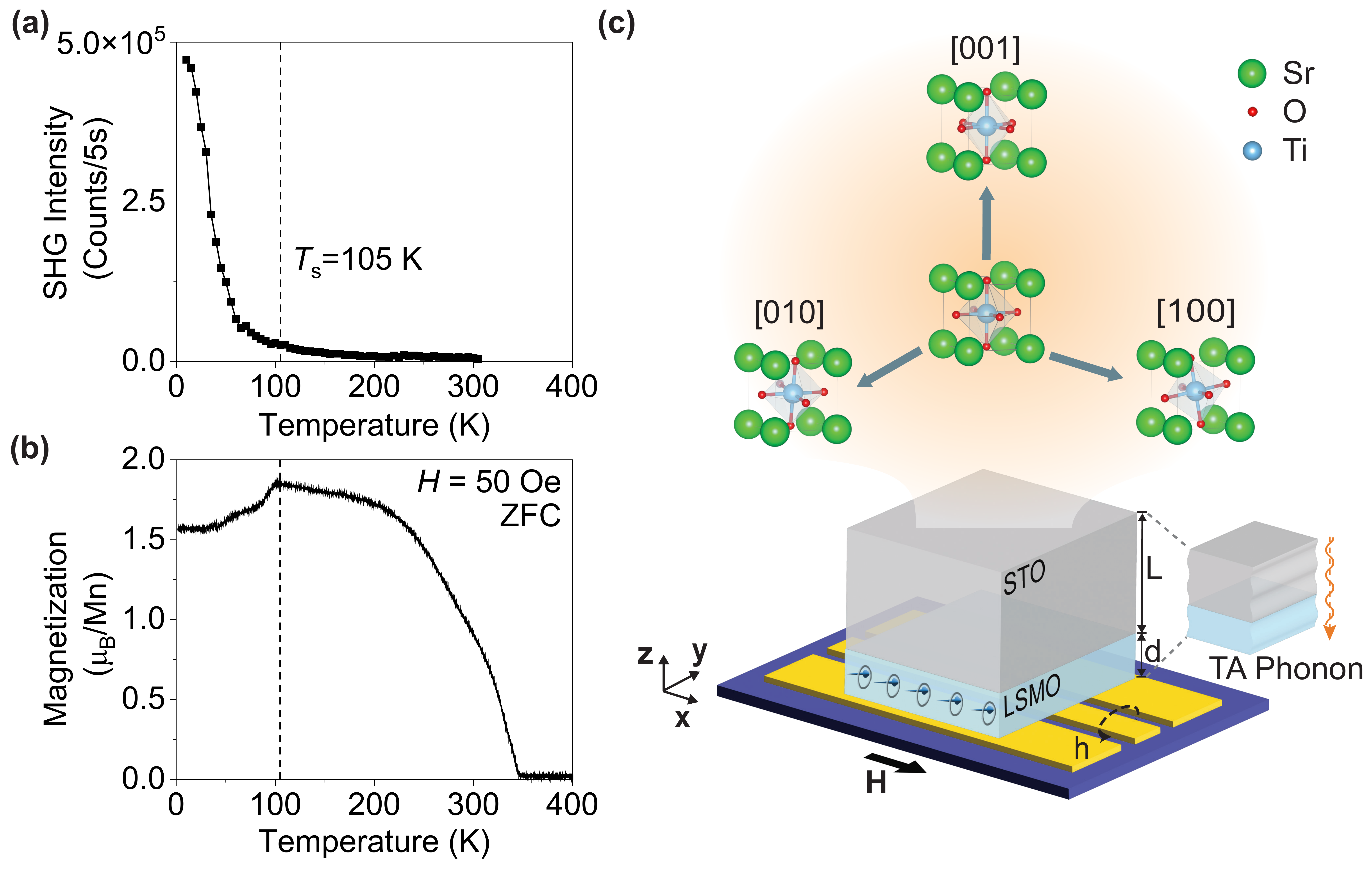}
\caption{\label{ModelDesign_V2} \textbf{(a)} Optical SHG intensity of \ce{LSMO}/\ce{SrTiO3} as a function of temperature acquired in normal incidence configuration. \textbf{(b)} In-plane magnetization as a function of temperature acquired in the zero-field-cool (ZFC) measurement with an external magnetic field of 50 Oe applied along the $x$-direction. Dashed lines in (a) \& (b) denote the cubic-to-tetragonal phase transition of STO at $T_\mathrm{S}$ = 105 K. \textbf{(c)} Schematic of the magnetoelastic heterostructure composed of LSMO thin films ($d$ = 38 nm) epitaxially grown on an STO substrate ($L$ = 0.5 mm). When the STO undergoes the cubic-to-tetragonal phase transition, the elongated axis can be along [100], [010], and [001] crystalline orientations. The sample is mounted on a coplanar waveguide for FMR measurements.}
\end{figure}

LSMO thin films (thickness, $d$ = 38 nm) were grown epitaxially on (001)-oriented STO substrates (thickness, $L$ = 0.5 mm) using pulsed laser deposition (see synthesis details in Methods). X-ray diffraction demonstrates excellent crystallinity and sharp interface of the LSMO/STO heterostructure (see Sec. 1 of the Supplemental Materials). Scanning transmission electron microscopy further confirms that the interface remains sharp after extensive cycles of cryogenic FMR measurements (see Sec. 2 of the Supplemental Materials). We first characterize the structural and magnetic properties of our sample. Figure \ref{ModelDesign_V2}(a) shows the intensity of optical second harmonic generation (SHG) of the LSMO/STO heterostructure. LSMO thin films preserve a centrosymmetric perovskite structure under tensile strain from the STO substrate. Meanwhile, STO is centrosymmetric in both cubic (O$_\mathrm{h}$ point group) and tetragonal (D$_\mathrm{4h}$ point group) phases. Therefore, none of these structures supports bulk electric dipole SHG. We thus attribute the pronounced onset and rapid growth of SHG intensity at $T_\mathrm{S}$ to symmetry breaking at the buried interface (see symmetry-resolved SHG analysis in Sec. 3 of the Supplemental Materials). Given that the cubic-to-tetragonal transition is driven by the rotation of the \ce{TiO6} octahedra about the lengthened axis (labeled as $c$ axis), as sketched in Fig.~\ref{ModelDesign_V2}(c), three types of tetragonal domains at micrometer-scale coexist with the lengthened axis oriented along the original cubic [100], [010], and [001] directions ~\cite{kalisky2013locally}, generating anisotropic local strains fields. The temperature dependence of magnetization $M-T$ of the LSMO/STO heterostructure was acquired in zero-field-cooled (ZFC) measurements. As shown in Fig.~\ref{ModelDesign_V2}(b), LSMO is ferromagnetic below Curie temperature $T_\mathrm{C}$ $\sim$ 340 K~\cite{PhysRevB.101.024408}. See $M$-$H$ hysteresis measurements of magnetic anisotropy, saturation magnetization, and coercivity in Sec. 4 of the Supplemental Materials. Notably, a kink of magnetization at 105 K coincides with the structural transition of STO~\cite{PhysRevB.93.104403}, confirming the magnetoelastic coupling between LSMO and STO~\cite{PhysRevB.79.104417, PhysRevB.84.184412}. From 105 K to 10 K, the magnetization is suppressed from 1.8 $\mu_B/$\ce{Mn} to 1.6 $\mu_B/$\ce{Mn}, which is consistent with the previous report~\cite{LSMO_Static_magnetization}.

\begin{figure*}[ht!]
\includegraphics[width=1.0\textwidth]{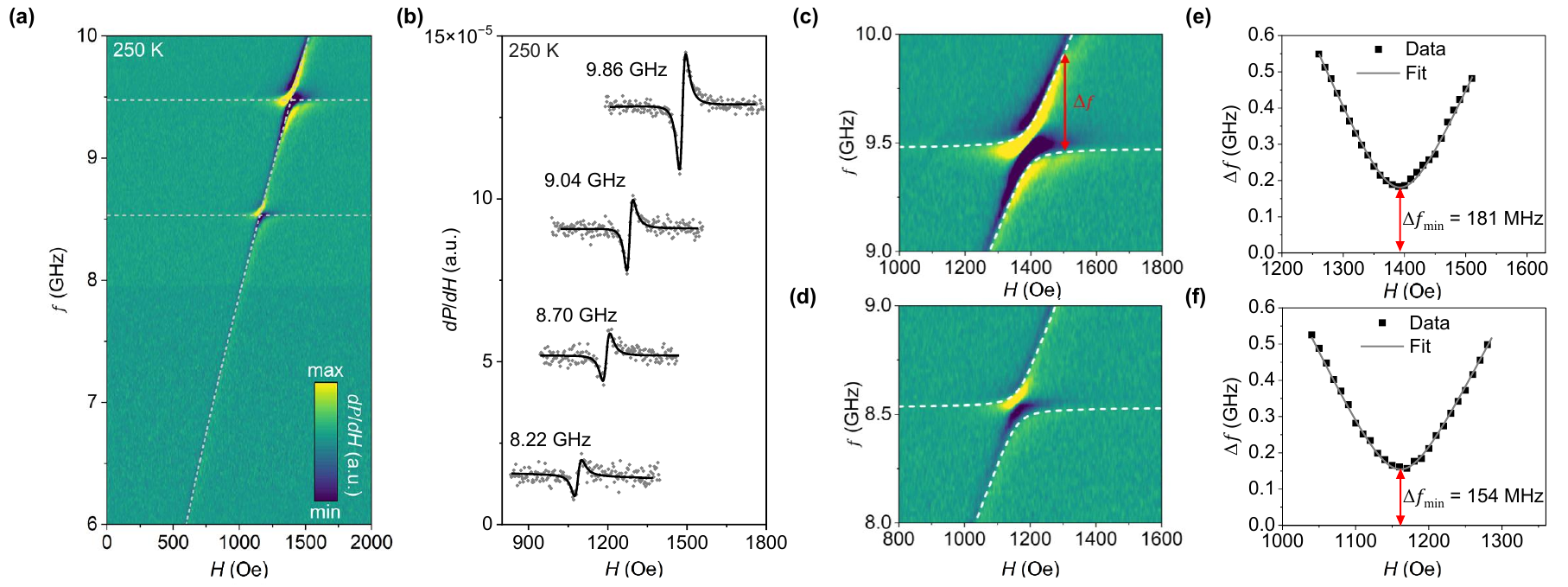}
\caption{\label{FMR250K} (a) Colormap of FMR spectral intensity of LSMO/STO heterostructure at 250 K. Dashed curve denotes the magnon mode and the horizontal dashed lines denote the phonon modes. (b) FMR line profiles at selected frequencies and corresponding fits to a superposition of symmetric and antisymmetric Lorentzian functions. FMR line profiles are vertically offset for clarity. Magnified view of avoided crossings at (c) $f_1$ = 9.47 GHz, $H_1=1391$ Oe, and (d) $f_2$ = 8.53 GHz, $H_2=1155$ Oe. The white dashed curves show the fits of the avoided crossings to the coupled magnon-phonon model. (e) and (f) show the corresponding splitting energy ($\Delta f$) for avoided crossings in (c) and (d), respectively. $\Delta f_\mathrm{min}$ represents the avoided crossing gap size.}
\end{figure*}

To examine the magnon-phonon coupling, we performed FMR measurements using the experimental geometry sketched in Fig.~\ref{ModelDesign_V2}(c). The LSMO/STO heterostructure was mounted facing down atop a coplanar waveguide through which the radio-frequency magnetic field ($h$) was injected to excite magnons via photon-magnon coupling, and subsequently excite phonons through magnon-phonon coupling. Unless otherwise specified, all FMR data are acquired with an external static magnetic field ($H$) along the [100] direction. Fig.~\ref{FMR250K}(a) shows the measured colormap of the field-modulated FMR spectral intensity ($dP/dH$) at 250 K, in which we identify a magnon band and two phonon bands. They intersect at $f_1$ = 9.47 GHz, $H_1=1391$ Oe, and $f_2$ = 8.53 GHz, $H_2=1155$ Oe, forming two avoided crossings. To quantify the magnon-phonon coupling strength, we first fit the FMR spectra using a superposition of symmetric and antisymmetric Lorentzian functions~\cite{tang2023spin}. Fitted FMR profiles at selected frequencies are shown in Fig.~\ref{FMR250K}(b). This analysis allows us to extract the resonance field and the spectral linewidth (see detailed fitting results in Sec. 5 of the Supplemental Materials). Figure~\ref{FMR250K}(c) shows the fitted avoided crossing associated with the $f_1$ = 9.47 GHz phonon mode using the coupled magnon-phonon model \cite{berk2019strongly}. The difference between the upper and lower branches ($\Delta f$) is shown in Fig.~\ref{FMR250K}(e), whose minimum value corresponds to the size of the avoided crossing gap ($\Delta\mathit{f}_\mathrm{min}$ = 181 MHz). Similar analyses of the avoided crossing at ($f_2$, $H_2$) are shown in Fig.~\ref{FMR250K}(d) and ~\ref{FMR250K}(f), yielding an avoided crossing gap of $\Delta\mathit{f}_\mathrm{min}$ = 154 MHz.

We use cooperativity, $C=\frac{\Gamma^2}{\kappa_\mathrm{m}\kappa_\mathrm{p}}$, to quantify the magnon-phonon coupling \cite{berk2019strongly}, where $\Gamma$ is the coupling strength given by the half of the avoided crossing gap size $\Gamma$ = $\Delta\mathit{f}_\mathrm{min}$/2. The $\kappa_\mathrm{m}$ and $\kappa_\mathrm{p}$ denote the dissipation rate (HWHM) of the magnon and phonon in the uncoupled regime, respectively. Here, the magnon dissipation rate can be estimated using $\kappa_\mathrm{m}$ = $\Delta f_0+\alpha\mathit{f}$, where $\Delta f_0$ is the inhomogeneous broadening, $\mathit{f}$ is the resonance frequency, and $\alpha$ is the effective damping constant~\cite{PhysRevB.109.014437}. The $\alpha$ value is extracted from the linear fit of the FMR linewidth (details in Sec. 5 of the Supplemental Materials). We obtain $\alpha$ = 0.003 $\pm$ 0.001 at 250 K, which is comparable to the previously reported value (0.0045) in LSMO thin films~\cite{acs.nanolett.4c02697}. Table I summarizes the coupling strength, dissipation rates, and cooperativity of the two magnon polarons, and both fall into the strong coupling regime ($C>1$)~\cite{PhysRevLett.113.156401}.

\begin{table}[h!]
\caption{Coupling strength ($\Gamma$), magnon dissipation rate ($\kappa_\mathrm{m}$), phonon
dissipation rate ($\kappa_\mathrm{p}$), and cooperativity ($C$) for the two
magnon-phonon avoided crossings at 250~K.}
\label{tab:ch5_coop}
\vspace{5pt}
\centering
\renewcommand{\arraystretch}{1.7}
\begin{tblr}{m{6.5cm} m{2.2cm} m{2.2cm} m{2.2cm} m{1.0cm}}
\hline
\textbf{Magnon-phonon avoided crossing} & \textbf{$\Gamma$ (MHz)} & \textbf{$\kappa_\mathrm{m}$ (MHz)}
& \textbf{$\kappa_\mathrm{p}$ (MHz)} & $C$ \\
\hline
$f_1 = 9.47$~GHz, $H_1=1391$ Oe & 91 & 67 & 65 & 1.87\\
$f_2 = 8.53$~GHz, $H_2=1155$ Oe & 77 & 64 & 39 & 2.37 \\
\hline
\end{tblr}
\end{table}

\begin{figure}[ht!]
\includegraphics[width=0.75\textwidth]{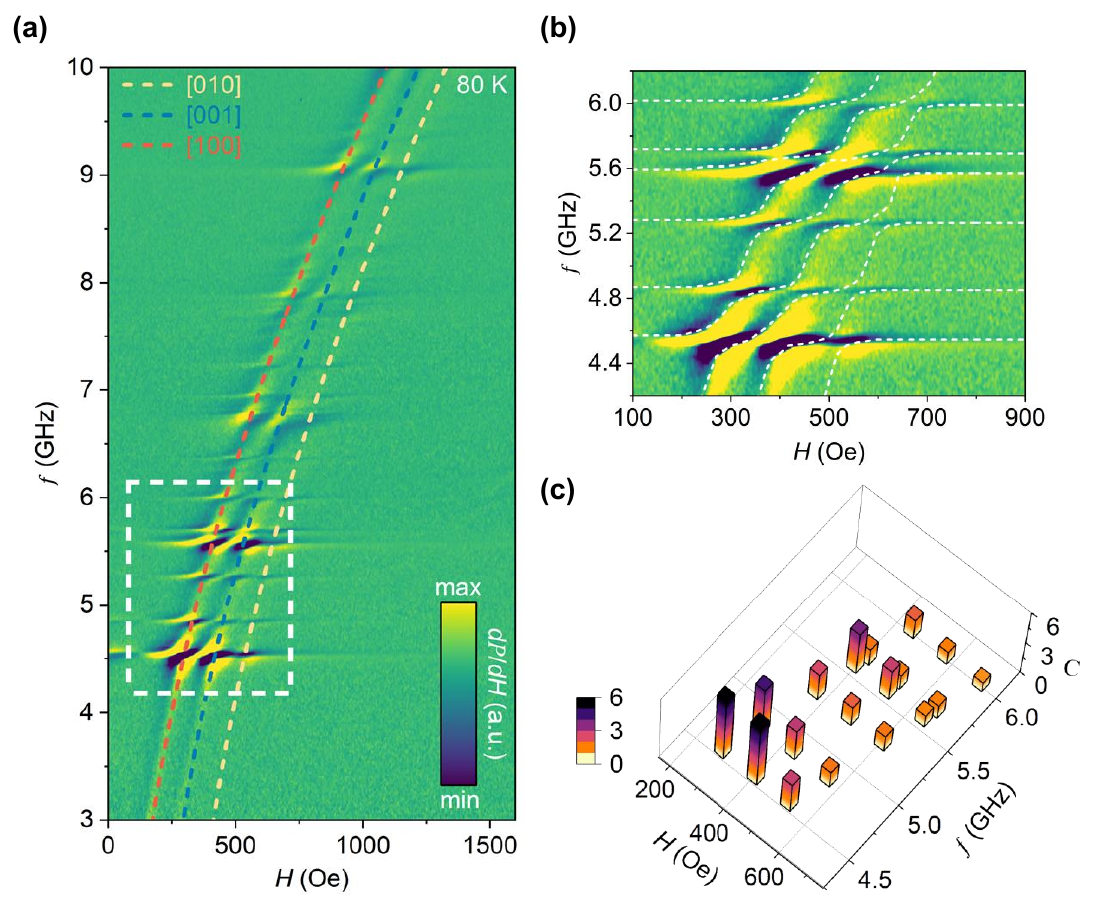}
\caption{\label{80K} \textbf{(a)} Colormap of FMR spectral intensity at 80 K. The local strain-driven split magnon bands are locked to the crystalline directions [100] (red), [010] (yellow), and [001] (blue), respectively. \textbf{(b)} Zoom-in of the region enclosed by the dashed box in (a). The white curves are guide-to-the-eye of the avoided crossings. \textbf{(c)} The cooperativity values of the avoided crossings from (b) as a function of frequency and magnetic field.}
\end{figure}

We then cool down the sample to $T$ = 80 K. As shown in Fig.~\ref{80K}(a), the most profound change is that the single magnon band splits into three branches. Upon further cooling to 50 K, the left and right magnon bands further separate from the middle one (see details in Sec. 6 of the Supplemental Material). Such splitting of the magnon band can be attributed to the structural phase transition of STO, in which the tetragonality breaks the isotropy of the crystalline axis of the cubic structure -- the three bands are associated with the elongated axis along [001], [100], and [010] directions, respectively. The assignment of the split magnon bands is rigorously corroborated by our analytical model, which will be discussed later. 

In addition, more phonon modes are present as a result of the reduced dissipation rate of phonons at low temperature. The periodic spacing of the phonon frequencies indicates that they are standing wave modes. Remarkably, all of them hybridize with the three magnon bands, forming a matrix of avoided crossings in the $f-H$ space. Figure~\ref{80K}(b) shows a magnified view of the avoided crossings enclosed in the white box in Fig.~\ref{80K}(a). We extract cooperativity values of these magnon polarons as shown in Fig.~\ref{80K}(c), and find that they span from 1.12 to 5.91. The matrix of avoided crossings resemble the characteristics of a hybrid magnon-phonon cavity in the multimode strong-coupling regime, which enables information to be encoded in and read out via frequency and magnetic field control.

To substantiate our experimental results, we develop an analytical model based on dynamic magnetoelastic boundary conditions~\cite{PhysRevB.104.014403}. In particular, we incorporate the effects of interfacial strain, which was not previously considered in similar phonon cavity models. We consider the LSMO/STO heterostructure as a 1D system where physical quantities only vary along the $z$ axis. The equation of motion that describes the mechanical displacement $u_{i} (z,t)$ and magnetization oscillation $m_{i} (z,t)$ ($i=x,y,z$) can be written as:

\begin{subequations} 
\begin{equation}
\label{eq:1a}
   \rho \frac{\partial ^{2}u_{x}}{\partial t^{2}}=
(1+\beta \frac{\partial }{\partial t})\left[c_{44}\frac{\partial ^{2}u_{x}}{\partial t^{2}} + {B_2}\frac{\partial (m_{x}m_{z})}{\partial z}\right]
\end{equation}

\begin{equation}
\label{eq:1b}
   \rho \frac{\partial ^{2}u_{y}}{\partial t^{2}}=
(1+\beta \frac{\partial }{\partial t})\left[c_{44}\frac{\partial ^{2}u_{y}}{\partial t^{2}} + {B_2}\frac{\partial (m_{y}m_{z})}{\partial z}\right]
\end{equation}

\begin{equation}
\label{eq:1c}
   \rho \frac{\partial ^{2}u_{z}}{\partial t^{2}}=
(1+\beta \frac{\partial }{\partial t})\left[c_{11}\frac{\partial ^{2}u_{z}}{\partial t^{2}} + {B_1}\frac{\partial (m_{z}^2)}{\partial z}\right]
\end{equation}
\end{subequations}

\noindent where $\rho$ and $\beta$ are the mass density and the viscous elastic damping coefficient; $c_{11}$ and $c_{44}$ are the elastic stiffness constants; and $B_1$ and $B_2$ are the magnetoelastic coupling coefficients (see detailed formalism in Sec. 7 of the Supplemental Materials). 

In the dynamical regime, we define mechanical displacement as $u_{i} (z,t) = u_{i}^{0}+\Delta u_{i}(z,t)$ and magnetization oscillation in the form of plane wave perturbation $\textbf{m} =\textbf{m}^0+\Delta \textbf{m}(t)=(m_{x}^{0}+\Delta m_{x}^{0}e^{-\rm{i}\omega t}, \Delta m_{y}^{0}e^{-\rm{i}\omega t}, \Delta m_{z}^{0}e^{-\rm{i}\omega t})$, where $\textbf{m}^0=(m_{x}^{0},m_{y}^{0},m_{z}^{0})=(1,0,0)$ when the external static magnetic field $H$ is applied along the $x$ axis. Then, we can derive the stress distribution $\sigma_{iz}(z,t)=\sigma_{iz}(t=0)+\Delta \sigma_{iz}(z,t)$ in LSMO thin film (superscript ‘m’):

\begin{subequations}    
\begin{equation}
\label{eq:2a}
   \Delta \sigma_{xz}^{m}(z,t)=c_{44}^{m} \frac{\partial\Delta u_{x}^{m}(z,t)}{\partial z}+B_{2}\Delta m_{z}^{0}e^{-i\omega t}
\end{equation}

\begin{equation}
\label{eq:2b}
   \Delta \sigma_{yz}^{m}(z,t)=c_{44}^{m} \frac{\partial\Delta u_{y}^{m}(z,t)}{\partial z}
\end{equation}

\begin{equation}
\label{eq:2c}
   \Delta \sigma_{zz}^{m}(z,t)=c_{11}^{m} \frac{\partial\Delta u_{z}^{m}(z,t)}{\partial z}
\end{equation}
\end{subequations}

Given that the GHz magnetic field ($h$) injected through the coplanar waveguide is expected to be spatially uniform in the LSMO thin film, it excites only the $k=0$ magnon (Kittel mode). As revealed from Eqs.(\ref{eq:2a})-(\ref{eq:2c}), the Kittel magnon $\Delta m_{z}=\Delta m_{z}^{0}e^{-\rm{i}\omega t}$ is only associated with $\Delta \sigma_{xz}^{m}$ through $B_2$. Therefore, excitation of the Kittel magnon only induces mechanical displacement that varies spatially along $z$ direction [$u_x(z,t)$], which corresponds to a TA phonon mode. We further demonstrate that this conclusion is valid in both the cubic phase ($T > T_\mathrm{S}$) and the tetragonal phase ($T < T_\mathrm{S}$) of the STO substrate (see Sec. 7 of the Supplemental Materials). 

Having identified the Kittel magnon and TA phonon involved in the magnetoelastic coupling, we note that the phonon with lower $k$ is expected to have better profile overlap with the Kittel magnon, while high-wavenumber magnon-phonon avoided crossing can be ignored in this work. Moreover, we assume that the propagating TA phonons obey the reflective boundary conditions imposed by the LSMO and STO surface and interface, and thus form standing waves. These standing waves can be considered as cavity modes whose resonance frequency and mode number can be expressed using the equation in Ref.~\cite{PhysRevB.104.014403}.

Based on the settings above, we carry out analytical simulations of the FMR absorption spectra $\textbf{P}_{\rm{abs}}(\omega, \textbf{H}^{\rm{bias}})$ using the Landau–Lifshitz–Gilbert equation, where $\omega$ is the microwave frequency and $\textbf{H}^{\rm{bias}}$ is the bias magnetic field applied along $x$ direction, i.e., $\textbf{H}^{\rm{bias}}=(H_{x}^{\rm{bias}},0,0)$ (see detailed methods in Sec. 8 of the Supplemental Materials). When STO is in the cubic phase ($T > T_{\mathrm{S}}$), the lattice mismatch strain between LSMO and STO is isotropic along the $x$ and $y$ axis, i.e., ${\varepsilon_{xx}} = {\varepsilon_{yy}} = \varepsilon^{\rm{mis}} = \frac{{{a_{{\rm{STO}}}} - {a_{{\rm{LSMO}}}}}}{{{a_{{\rm{LSMO}}}}}}=0.826\%$, where $a_{{\rm{STO}}} = 3.905 \text{\AA}$ and $a_{{\rm{LSMO}}} = 3.873 \text{\AA}$ are the lattice parameters of cubic STO substrate and LSMO thin film, respectively. In this case, the Kittel magnon frequency ($\omega_{\mathrm{m}}^{\rm{cubic}}$) can be expressed as 

\begin{eqnarray}
    \label{eq:3}
    \begin{aligned}
       \omega_{\rm{m}}^{\rm{cubic}} = \frac{\gamma }{{{\mu _0}{M_s}}}\sqrt {\frac{ {A_0} + A_c}{{3{c_{11}}{c_{44}}}}} \sqrt { ({{\mu _0}{M_s}H_{x}^{\rm{bias}} + 2{K_1}})} 
    \end{aligned}
\end{eqnarray}

\noindent where $\gamma$ is the gyromagnetic ratio, $M_s$ is the saturation magnetization, and ${A_0} = - 3{c_{11}}B_2^2 + 2{c_{44}}B_1^2 + 3{c_{11}}{c_{44}}\left( {2{K_1} + {\mu _0}{M_s}\left( {H_{x}^{\rm{bias}} + {M_s}} \right)} \right)$. $A_c = -6{B_1}{c_{44}}\left( {{c_{11}} + 2{c_{12}}} \right){\varepsilon^{\rm{mis}}}$ is related to the lattice mismatch strain (see Sec. 9 of the Supplemental Materials). As shown in Fig.~\ref{Model}(a), we obtain the simulated FMR colormap at 250 K using the parameters summarized in Sec. 10 of the Supplemental Materials. 

When STO is in the tetragonal phase ($T < T_{\rm{S}}$), anisotropic interfacial strains are produced as ${\varepsilon _{a}} = \frac{{{a_{{\rm{STO}}}} - {a_{{\rm{LSMO}}}}}}{{{a_{{\rm{LSMO}}}}}} $ does not equal to ${\varepsilon _{c}} = \frac{{{c_{{\rm{STO}}}} - {a_{{\rm{LSMO}}}}}}{{{a_{{\rm{LSMO}}}}}} $ ($a_{{\rm{STO}}}$ and $c_{{\rm{STO}}}$ are the lattice parameters of the tetragonal STO). Since the lengthened $c_{{\rm{STO}}}$ axis can be along the [100], [010], and [001] directions, we obtain three bands of Kittel magnon as expressed below:

\begin{subequations}

\begin{eqnarray}
    \label{eq:4a}
    \begin{aligned}
       \omega _{\rm{m}}^{\left[ {001} \right]} = \frac{\gamma }{{{\mu _0}{M_s}}}\sqrt {\frac{{A_0}+A_{[001]}}{{3{c_{11}}{c_{44}}}}} \sqrt { ({{\mu _0}{M_s}H_{x}^{\rm{bias}} + 2{K_1}})} 
    \end{aligned}
\end{eqnarray}

\begin{eqnarray}
    \label{eq:4b}
    \begin{aligned}
        \omega _{\rm{m}}^{\left[ {100} \right]} &= \frac{\gamma }{{{\mu _0}{M_s}}}\sqrt {\frac{{A_0}+{A_{[100]}}}{{3{c_{11}}{c_{44}}}}} \sqrt {\left( {2{B_1}\left( {{\varepsilon _c} - {\varepsilon _a}} \right) + {\mu _0}{M_s}H_{x}^{\rm{bias}} + 2{K_1}} \right)} 
    \end{aligned}
\end{eqnarray}

\begin{eqnarray}
    \label{eq:4c}
    \begin{aligned}
        \omega _{\rm{m}}^{\left[ {010} \right]} &= \frac{\gamma }{{{\mu _0}{M_s}}}\sqrt {\frac{{A_0}+A_{[010]}}{{3{c_{11}}{c_{44}}}}}\sqrt {\left( {2{B_1}\left( {{\varepsilon _a} - {\varepsilon _c}} \right) + {\mu _0}{M_s}H_{x}^{\rm{bias}} + 2{K_1}} \right) } 
    \end{aligned}
\end{eqnarray}

\end{subequations}

\noindent where $A_{[001]}=-6{B_1}{c_{44}}({{c_{11}} + 2{c_{12}}}){\varepsilon_c}$, $A_{[100]}=-6{B_1}{c_{44}}[({{c_{11}} + {c_{12}}}){\varepsilon_a}+{c_{12}}{\varepsilon_c}]$, and $A_{[010]}=-6{B_1}{c_{44}}[({{c_{11}} + {c_{12}}}){\varepsilon_c}+{c_{12}}{\varepsilon _a}]$ are numerically very close to $A_c$ of the cubic phase. 

By comparing the Eq.(\ref{eq:3}) and Eq.(\ref{eq:4a}), we note that $\omega_\text{m}^{[001]}$ remains nearly identical to $\omega_\text{m}^\text{cubic}$ due to the subtle difference between ${\varepsilon}^{\rm{mis}}$ and $\varepsilon_c$. In contrast, as reflected in Eq.(\ref{eq:4b}) and Eq.(\ref{eq:4c}), the $2B_1(\varepsilon_c-\varepsilon_a)$ term in $\omega_\text{m}^{[100]}$ and the $2B_1(\varepsilon_a-\varepsilon_c)$ term in $\omega_\text{m}^{[010]}$ cause these magnon bands to shift toward the lower and higher magnetic field directions, respectively. Therefore, our model clearly justifies our assignment of split magnon bands locked to [100], [010], and [001] crystalline orientations in Fig.~\ref{80K}(a). Quantitatively, our simulation reproduces the enhanced magnon splitting from 80 K (Fig.~\ref{Model}(b)) to 50 K (Fig.~\ref{Model}(c)), when the interfacial strain changes from ${\varepsilon _{c}} = 1.008\% $ and ${\varepsilon _{a}} = 0.976\%$ to ${\varepsilon _{c}} = 1.078\% $ and ${\varepsilon _{a}} = 0.946\%$. Note that the anisotropic strain ($\varepsilon _{c}-\varepsilon _{a}$) emerged from the cubic-to-tetragonal phase transition is only 0.032\% at 80 K. Such precise strain engineering is mainly benefit from the large magnetoelastic coupling constants of the LSMO (see $B_1$, $B_2$ values in Sec. 10 of the Supplemental Materials), which are comparable with those of Ni, Fe, and Co, and larger than those of YIG~\cite{li2021advances}.

Finally, we quantify the magnon-phonon coupling strength at 250 K (see Sec. 11 of the Supplemental Materials). Figure~\ref{Model}(d) shows the coupling strength ($\Gamma$) as a function of the frequency of standing wave phonons, which reproduces our measured avoided crossing gaps at 8.53 GHz and 9.47 GHz (see Fig.~\ref{FMR250K}(c)-\ref{FMR250K}(f)). Compared with the analytical expression for magnetoelastic coupling strength in Ref.~\cite{PhysRevB.101.060407}, our model accounts for more material parameters such as anisotropy coefficient, elastic stiffness coefficients, and lattice-mismatch strain. Moreover, our model can be properly applied to the cases when $B_1$ and/or $B_2$ are negative.      

In summary, we have demonstrated that LSMO/STO magnetoelastic heterostructures exhibit strong magnon-phonon coupling. When the STO substrate undergoes a transition into the tetragonal phase, the Kittel magnon in LSMO splits into three bands due to the anisotropic interfacial strain, forming a matrix of hybrid magnon-phonon modes. This local strain-driven hybrid magnon-phonon cavity has far-reaching implications beyond the prototypical hybrid systems based on magnetic multilayers. First, it allows precise strain engineering by growing LSMO thin films on various lattice-mismatched substrates, or freestanding STO nanomembrane ~\cite{xu.10.1038, zhu2025theory}. Second, our current analytical model can capture the key physical principles very well, yet we envision a mesoscale model that can implement the complex structural domain patterns ~\cite{li2006phase} will facilitate future design of such hybrid systems. Finally, changes in temperature is employed to trigger the structural phase transition in this work, while we highlight that additional tuning knobs, such as an external electric field or mechanical stress, can be applied by integrating the LSMO film into a piezostage or piezoelectric underlayer~\cite{Nat.Comm.10.1038,hong2020extreme}. In a broader sense, by growing magnetic thin films epitaxially on a ferroelectric/ferroelastic underlayer, similar local strain fields can be produced isothermally and electrically by voltage-driven domain switching, opening up the potential of realizing on-chip, voltage-controlled multimode magnon-phonon hybridization with ultralow power consumption.\\

\section{Methods}
\vspace{-10pt}
\noindent\textbf{Synthesis:} LSMO films (thickness, $d$ = 38 nm) were grown epitaxially on a (001)-oriented \ce{SrTiO3} substrates (thickness, $L$ = 0.5 mm) using pulsed laser deposition with a laser fluence of approximately 1 J$\cdot$cm$^{-2}$ and frequency of 1 Hz. During the growth, the substrate temperature was held at 700 °C with an oxygen pressure of 0.3 Torr. The film was slowly cooled to room temperature postdeposition in 300 Torr \ce{O2} to ensure proper oxygen stoichiometry.

\vspace{5pt}
\noindent\textbf{X-ray Diffraction (XRD):} As-grown LSMO/STO samples were characterized by high-resolution XRD using a Rigaku SmartLab diffractometer. Rocking curves ($\omega-2\theta$ scans) were recorded and fitted to the XRD profile simulation using LEPTOS software.

\vspace{5pt}
\noindent \textbf{Physical Property Measurement System (PPMS):}  The static magnetization and ferromagnetic resonance spectroscopy were measured using a Quantum Design DynaCool PPMS (1.8 K and 9 T). The $M-T$ curves and $M-H$ hysteresis loops were acquired using the vibrating sample magnetometer option. The FMR measurements were performed using a coplanar waveguide (CPW) to inject the microwave, and an RF diode to convert the current signal into a detectable voltage. The sample was mounted face down in contact with the CPW. The FMR spectroscopy was acquired using the broadband CryoFMR option in field-modulated mode, in which an AC signal was supplied via the Helmholtz coils to generate a modulation perpendicular to the applied external field, and the output signal was recorded by a locked-in amplifier.

\vspace{5pt}
\noindent \textbf{Optical Second Harmonic Generation (SHG):} Optical SHG measurements were performed using normal incidence geometry. The intensity of the reflected SHG is recorded as a function of the azimuthal angle $\phi$ between the scattering plane and the in-plane crystalline axis. The incident ultrafast light source had 800 nm wavelength, 50 fs pulse duration, and 200 kHz repetition rate. The laser beam was focused into a 50 $\mu$m diameter spot on the sample with an average power of 1 mW. The intensity of the reflected SHG was measured using a single-photon counting detector. All thermal cycles were carried out in a Montana Instruments closed-cycle cryostation with a base pressure better than $5\times10^{-7}$ mbar.

\vspace{5pt}
\noindent\textbf{Scanning Transmission Electron Microscopy (STEM):} LSMO/STO samples were imaged via scanning electron microscopy and prepared for STEM analysis using a ThermoFisher Hydra dual-beam plasma focused ion beam using a standard liftout procedure. The samples were thinned to electron transparency with Xe ions at 30 kV down to 5 kV accelerating voltage with a final polish at 2 kV. STEM imaging and chemical mapping by energy dispersive x-ray spectroscopy were conducted on a JOEL GrandARM at 300 kV. Atomic resolution images were collected for geometric phase analysis (GPA) strain mapping with horizontal field widths of 24.5 and 37.5 nm (8 Mx and 5 Mx magnification, respectively). GPA was conducted using the application Strain++ version 1.8 (as maintained on GitHub by J. J. P. Peters https://jjppeters.github.io/Strainpp/) based on the first-order [001] and [010] fast Fourier transformation peaks.

\vspace{-5pt}
\section{Data availability}
\vspace{-15pt}
\noindent All data that supports the plots within this paper and other findings of this study are available from the corresponding author upon reasonable request.

\vspace{-5pt}
\section{Acknowledgments}
\vspace{-15pt}
\noindent We acknowledge the useful discussions with Chunhui Du, Peng Li, Ryan Comes, Valeria Lauter, Gwo Ching Wang, and Frank Y. Gao. PPMS measurements were carried out using the resources of the Alabama Micro/Nano Science and Technology Center (AMNSTC) at Auburn University. W.J. acknowledges support from the Air Force Office of Scientific Research under Grant No. FA9550-23-1-0499 and the NSF CAREER Award under Grant No. DMR-2339615. D.S. and Y.T. are supported by the NSF under Grant No. DMR-1745450. This research used a Rigaku SmartLab thin film x-ray diffractometer in the UC Davis Advanced Materials Characterization and Testing (AMCaT) Laboratory, acquired with funding from the NSF Major Research Instrumentation (MRI) Program under Award No. MRI-2216198. The work at UNC-CH is supported by the U.S. Department of Energy, Office of Science, Basic Energy Sciences under Award No. DE-SC0026305 (W.Z.). The work at the University of Wisconsin-Madison was primarily supported by the U.S. Department of Energy, Office of Science, Basic Energy Sciences, under Award No. DE-SC0020145 as part of the Computational Materials Sciences Program (Y. Z. and J.-M. H.), and partially supported by the Wisconsin MRSEC (DMR-2309000) (J. W. and J.-M. H.) STEM measurements and data analysis at PNNL were supported by the US Department of Energy, Office of Science, Basic Energy Sciences, Division of Materials Sciences and Engineering, Synthesis and Processing Science Program, under Award No. 10122.

\vspace{-5pt}
\section{Author contributions}
\vspace{-15pt}
\noindent W.J. and W.Z. conceived the project; D.S. and Y.T. synthesized the LSMO/STO sample and performed the XRD characterization; C.T. performed the FMR and VSM measurements with technical assistance of H.G. under the guidance of W.J., M.M.S., and M.A.; C.T. performed the optical SHG measurements under the guidance of W.J. and X.M.; C.T., Y.X., W.Z. and W.J. analyzed the data; Y.Z., J.W., and J.-M.H. developed the analytical model; B.E.M., L.W., and Y.D. carried out the STEM measurements; W.J., W.Z., and J.-M.H. wrote the manuscript. All authors discussed the results and contributed to the preparation of the manuscript.

\vspace{-5pt}
\section{Competing interests}
\vspace{-15pt}
\noindent The authors declare no competing interests.

\vspace{-5pt}
\section{Data availability}
\vspace{-15pt}
\noindent The datasets generated and/or analyzed during the current study are available from the corresponding author on reasonable request.\\

\bibliographystyle{naturemag}
\nocite{apsrev41Control}
\bibliography{references.bib}


\newpage
\textbf{Figure 1}\\

\newpage
\textbf{Figure 2}\\

\newpage
\textbf{Figure 3}\\

\newpage
\textbf{Figure 4}\\

\begin{figure}[ht!]
\includegraphics[width=0.75\textwidth]{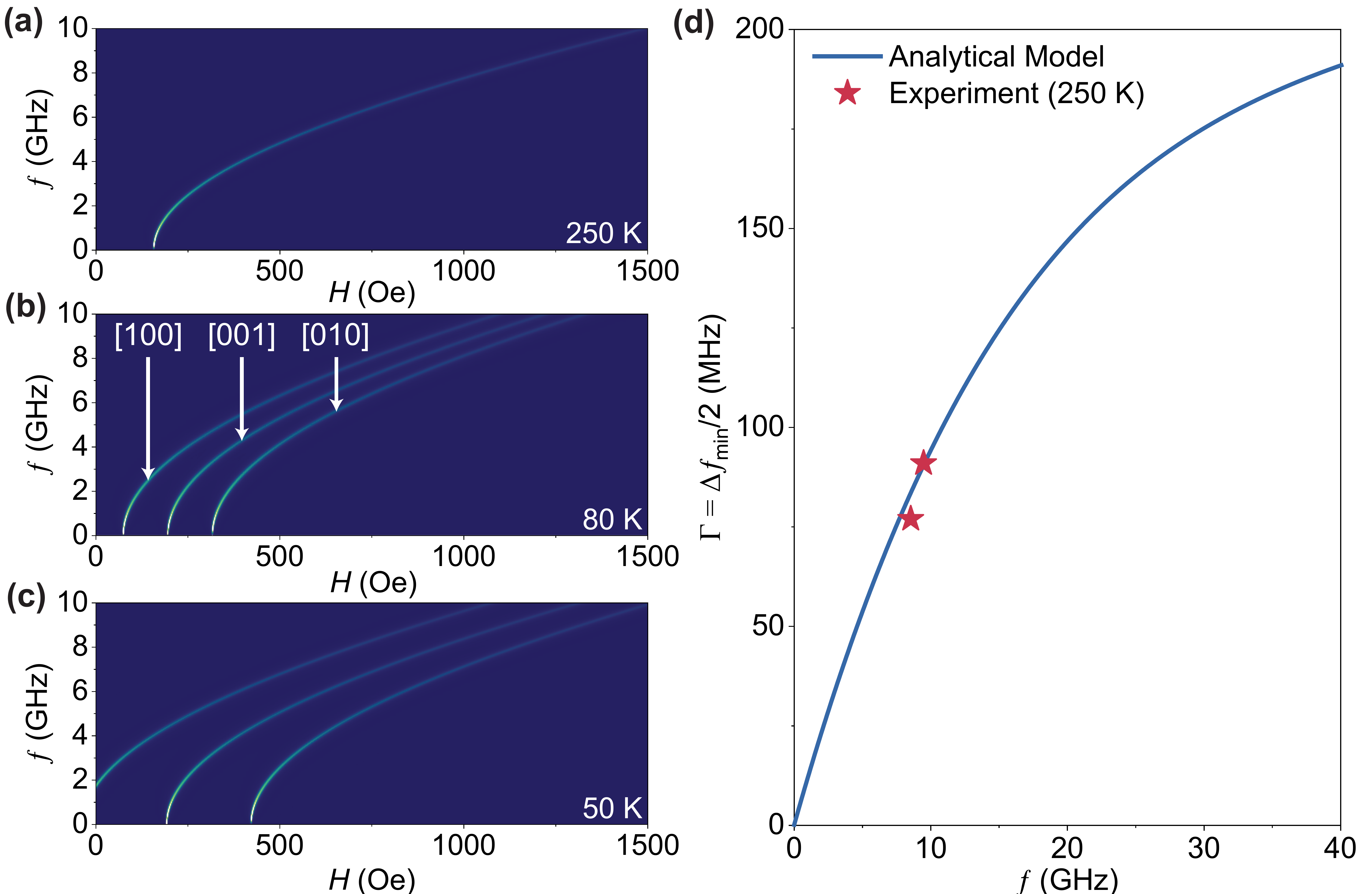}
\caption{\label{Model} \textbf{(a)} Simulated FMR of magnon band at 250 K derived from Eq.(3), where STO is in the cubic phase. Simulated FMR of split magnon bands associated with the [100], [010], and [001] crystalline orientations at \textbf{(b)} 80 K and \textbf{(c)} 50 K derived from Eq.(4a)-(4c), where STO is in the tetragonal phase. \textbf{(d)} Magnon-phonon coupling strength $\Gamma = \Delta f_{min}/2$ derived from analytical model Eq.(S37) in Sec.11 of Supplemental Materials. The measured $\Gamma$ values at 250 K (red stars) are overlaid for comparison.}
\end{figure}

\end{document}


\preprint{}

\title{Supplemental Materials \\ \ \\ 
\normalsize Multimode magnon-phonon cavity driven by symmetry-locked strain fields}

\author{Chunli Tang}
\altaffiliation{These two authors contributed equally}
\affiliation{Department of Electrical and Computer Engineering, Auburn University, Auburn, AL 36849, USA}
\affiliation{Department of Physics, Auburn University, Auburn, AL 36849, USA}

\author{Yujie Zhu}
\altaffiliation{These two authors contributed equally}
\affiliation{Department of Materials Science and Engineering, University of Wisconsin-Madison, Madison, WI 53706, USA}

\author{Dayne Sasaki}
\affiliation{Department of Materials Science and Engineering, University of California, Davis, Davis, CA 95616, USA}

\author{Jiaxuan Wu}
\affiliation{Department of Materials Science and Engineering, University of Wisconsin-Madison, Madison, WI 53706, USA}

\author{Yuzan Xiong}
\affiliation{Department of Physics and Astronomy, University of North Carolina at Chapel Hill, Chapel Hill, NC 27599, USA}

\author{Harshil Goyal}
\affiliation{Department of Electrical and Computer Engineering, Auburn University, Auburn, AL 36849, USA}
\affiliation{Department of Physics, Auburn University, Auburn, AL 36849, USA}

\author{Masoud Mahjouri-Samani}
\affiliation{Department of Electrical and Computer Engineering, Auburn University, Auburn, AL 36849, USA}

\author{Mark Adams}
\affiliation{Department of Electrical and Computer Engineering, Auburn University, Auburn, AL 36849, USA}

\author{Xiang Meng}
\affiliation{Department of Electrical Engineering, Columbia University, New York, NY 10027, USA}

\author{Bethany E. Matthews}
\affiliation{Environmental Molecular Sciences Laboratory, Pacific Northwest National Laboratory, Richland, Washington 99354, USA}

\author{Le Wang}
\affiliation{Environmental Molecular Sciences Laboratory, Pacific Northwest National Laboratory, Richland, Washington 99354, USA}

\author{Yingge Du}
\affiliation{Environmental Molecular Sciences Laboratory, Pacific Northwest National Laboratory, Richland, Washington 99354, USA}

\author{Jia-Mian Hu}
\affiliation{Department of Materials Science and Engineering, University of Wisconsin-Madison, Madison, WI 53706, USA}

\author{Yayoi Takamura}
\affiliation{Department of Materials Science and Engineering, University of California, Davis, Davis, CA 95616, USA}

\author{Wei Zhang}
\affiliation{Department of Physics and Astronomy, University of North Carolina at Chapel Hill, Chapel Hill, NC 27599, USA}

\author{Wencan Jin}
\email{wjin@auburn.edu}
\affiliation{Department of Physics, Auburn University, Auburn, AL 36849, USA}
\affiliation{Department of Electrical and Computer Engineering, Auburn University, Auburn, AL 36849, USA} 

\maketitle
\tableofcontents

\newpage
\section{$\textbf{S1. Sharp LSMO/STO interface characterized by XRD after growth}$}

X-ray diffraction (XRD) measurements were carried out using a Rigaku SmartLab diffractometer. As shown in \textbf{Fig. S1}, the (002) planes of the LSMO thin films are parallel to the family of $(00l)$ planes of the STO substrate. The clear Pendellosung fringes indicated the high crystallinity of the film and sharp interface. XRD fits were obtained using LEPTOS software, yielding the thickness of the LSMO thin films of 38 nm.

\begin{figure*}[ht!]
\includegraphics[width=0.5\textwidth]{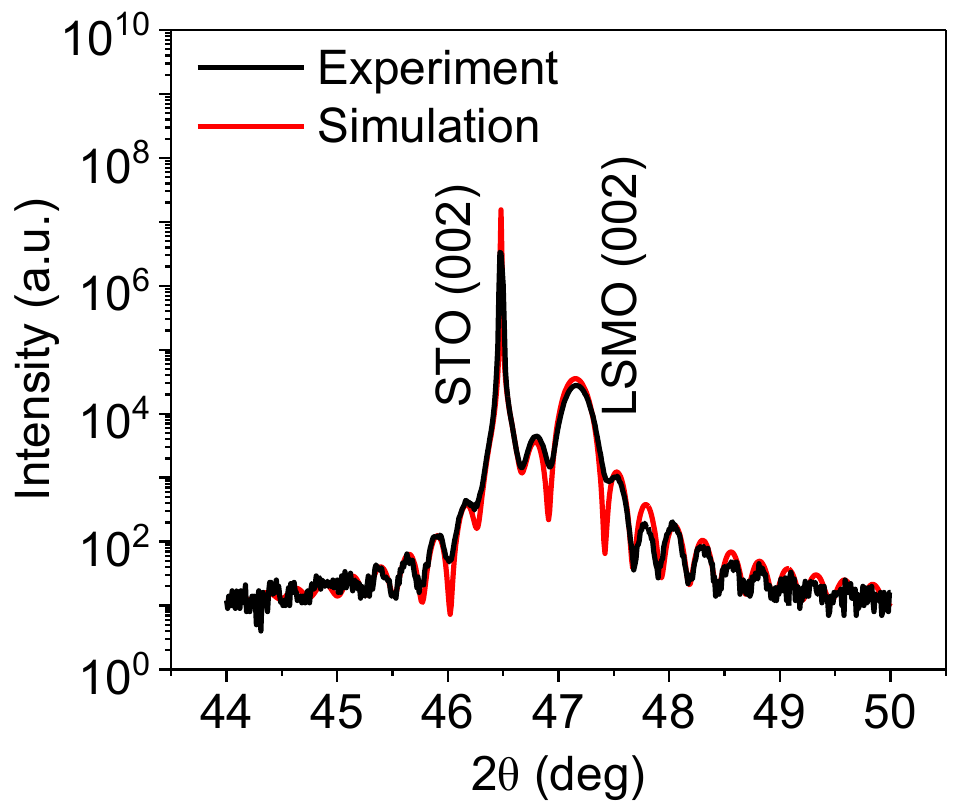}
\caption{\label{XRD} \justifying \small Experimental (black curve) and simulated (red curve) XRD scan around the (002) Bragg reflection of LSMO/STO heterostructure. }
\end{figure*}


\newpage
\section{$\textbf{S2. STEM characterization of interface and geometric phase analysis}$}

Scanning transmission electron microscopy (STEM) measurements were carried out at room temperature. As shown in \textbf{Fig. S2(a)}, high-angle annular dark-field (HAADF) imaging revealed the LSMO films to be primarily uniform and smooth, single crystal, with a thickness of 38 $\pm$ 0.5 nm, and epitaxially grown on STO. Occasional Mn-rich precipitates cluster near the substrate (see \textbf{Fig. S2(c)}), often associated with Sr/La compositionally non-uniform regions which appear directly above the Mn-rich regions. The corresponding strain mapping using geometric phase analysis (GPA) indicates no in-plane strain at the defect-free interface (see \textbf{Fig. S2(b)}) and a very slight increase in in-plane strain at the Mn-rich region (see \textbf{Fig. S2(d)}). Therefore, we confirm that the interface of the as-grown sample has negligible in-plane strain, and the magnon splitting is solely driven by the symmetry-locked strain field from the structural phase transition. 

\begin{figure*}[ht!]
\includegraphics[width=1.0\textwidth]{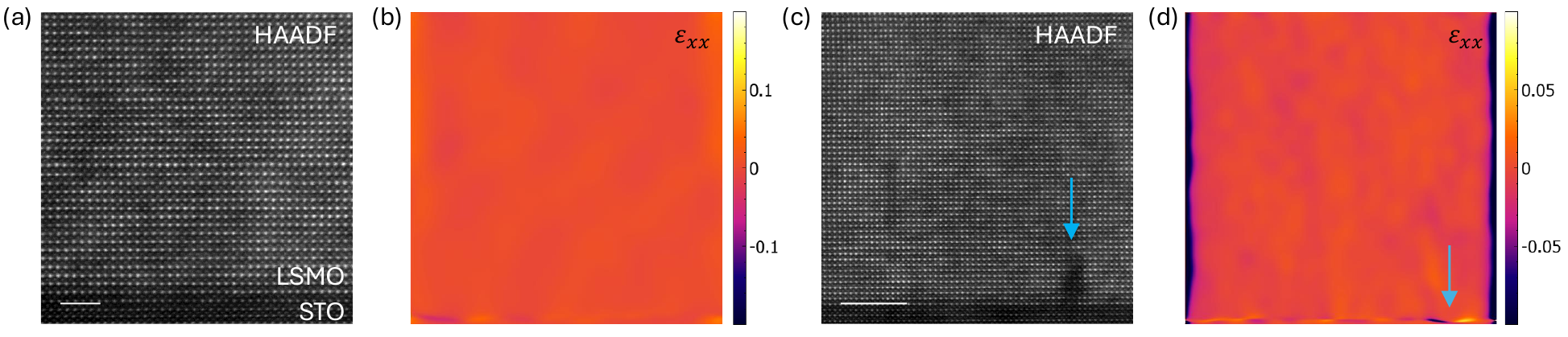}
\caption{\label{XRD} \justifying \small \textbf{(a)} HAADF image of the LSMO/STO interface. The scale bar is 2 nm. \textbf{(b)} GPA image corresponds to (a) shows no in-plane strain at the interface. \textbf{(c)} HAADF image of the LSMO/STO interface with a Mn-rich precipitates cluster marked by the blue arrow. The scale bar is 5 nm. \textbf{(d)} GPA image corresponds to (c) shows a very slight increase in in-plane strain at the Mn-rich region.}
\end{figure*}


\newpage
\section{$\textbf{S3. Cubic-to-Tetragonal structural phase transition of \ce{SrTiO3}}$ }

STO is in the centrosymmetric cubic phase (O$_\mathrm{h}$ point group) at high temperature ($T > T_\mathrm{S}$), and in the centrosymmetric tetragonal phase (D$_\mathrm{4h}$ point group) at low temperature ($T < T_\mathrm{S}$). These point groups do not support bulk electric dipole (ED) SHG through the full temperature range. Therefore, we consider the symmetry breaking at the interface. The bulk ED-SHG can be expressed as 

\begin{equation}
\label{eq:S1}
    P_{i}^{ED}(2\omega)=\sum_{jk}\chi_{ijk}^{(2)}E_j(\omega)E_k(\omega)\\
\end{equation}

\noindent 1) For the high temperature cubic phase, the (001) surface is in noncentrosymmetric C$_\mathrm{4v}$ point group, which allows 7 nonzero elements of $\chi_{ijk}^{(2)}$, and among them only 4 are independent:

\begin{center}
   xzx = yzy, xxz = yyz, zxx = zyy, zzz\\
\end{center}

\noindent One can write the susceptibility tensor as

\begin{equation}
\label{eq:S2}
\mathbf{\chi}_{ijk}^{(2)}(C_\mathrm{4v}) =
\begin{pmatrix}
\begin{bmatrix} 0 \\ 0 \\ \chi_{xxz} \end{bmatrix} &
\begin{bmatrix} 0 \\ 0 \\ 0 \end{bmatrix} &
\begin{bmatrix} \chi_{xxz} \\ 0 \\ 0 \end{bmatrix} \\[3.5em]

\begin{bmatrix} 0 \\ 0 \\ 0 \end{bmatrix} &
\begin{bmatrix} 0 \\ 0 \\ \chi_{xxz} \end{bmatrix} &
\begin{bmatrix} 0 \\ \chi_{xxz} \\ 0 \end{bmatrix} \\[3.5em]

\begin{bmatrix} \chi_{zxx} \\ 0 \\ 0 \end{bmatrix} &
\begin{bmatrix} 0 \\ \chi_{zxx} \\ 0 \end{bmatrix} &
\begin{bmatrix} 0 \\ 0 \\ \chi_{zzz} \end{bmatrix}
\end{pmatrix}
\end{equation}

\vspace{12pt}
\noindent In normal incidence geometry, ED-SHG vanishes in both parallel and crossed polarization configurations.\\

\begin{figure*}[ht!]
\includegraphics[width=1.0\textwidth]{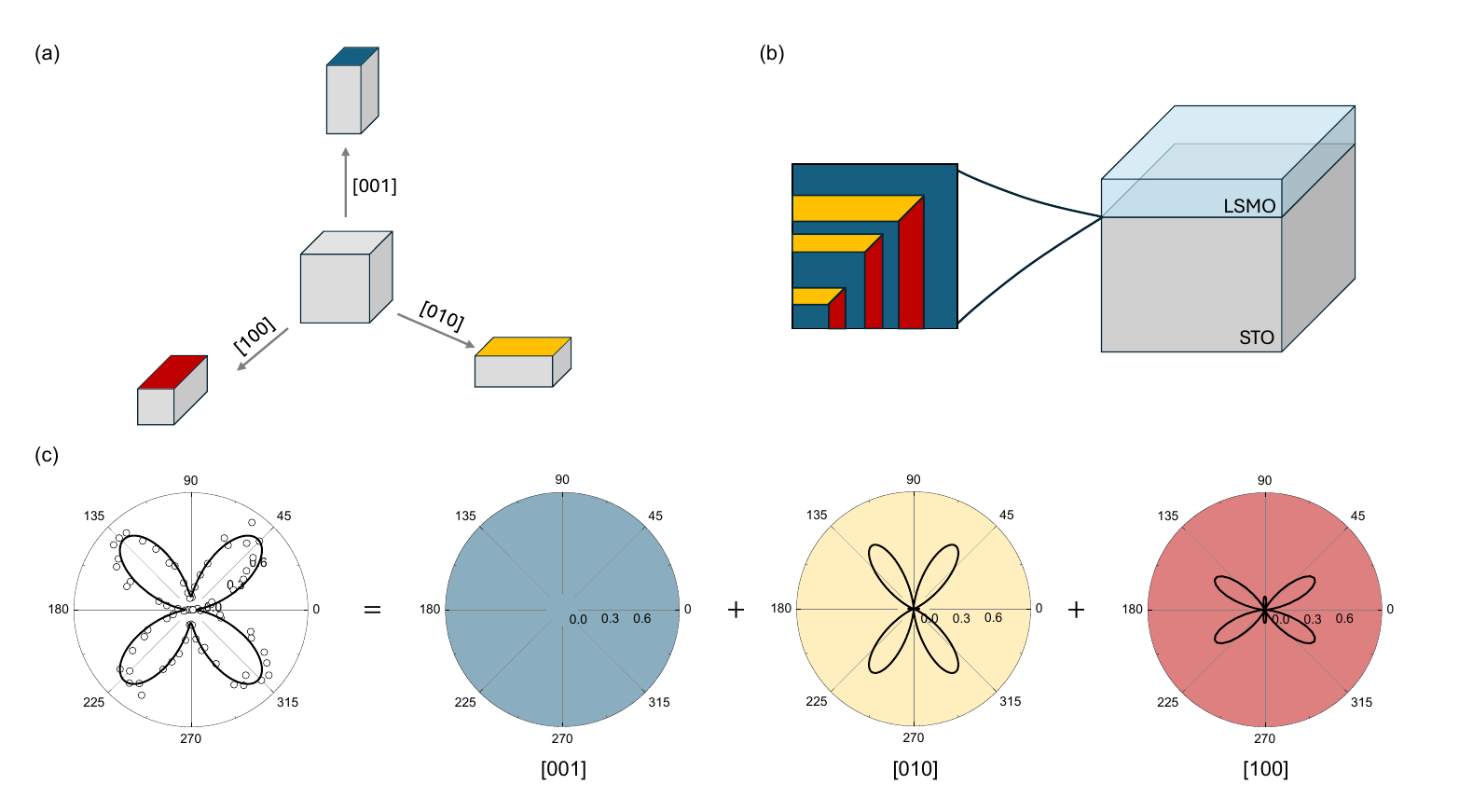}
\caption{\label{SHG} \justifying \small \textbf{(a)} The tetragonal phase breaks the cubic symmetry of STO and form three types of domains with the lengthened axis oriented along the original cubic [100], [010], and [001] directions. \textbf{(b)} The LSMO/STO interface consists of surfaces from the three types domains. Surfaces of the [100], [010], and [001] domains are denoted in red, yellow, and blue, respectively. \textbf{(c)} RA-SHG pattern of LSMO/STO interface acquired at 10 K and the fit to the superposition of ED-SHG from the surfaces of [100] (red), [010] (yellow), and [001] (blue) domains.}
\end{figure*}

\noindent 2) For the low temperature tetragonal phase of STO, as sketched in \textbf{Fig.~\ref{SHG}(a)}, there exist three types of domains with the lengthened axis oriented along the original cubic [100], [010], and [001] directions. The surface of the [001] domain (blue) is in C$_\mathrm{4v}$ point group, while the surfaces of the [100] domain (red) and the [010] domain (yellow) are in C$_\mathrm{2v}$ point group. Therefore, as sketched in \textbf{Fig.~\ref{SHG}(b)}, the LSMO/STO interface at low temperature consists of these three types of the tetragonal domains with equal population.

As derived above, the surface of the [001] domain (blue) in C$_\mathrm{4v}$ point group has no contribution to the ED-SHG in normal incidence geometry. Below we derive the ED-SHG contribution from the surfaces of the [100] domain (red) and the [010] domain (yellow) in C$_\mathrm{2v}$ point group. The noncentrosymmetric C$_\mathrm{2v}$ point group allows 7 nonzero elements of $\chi_{ijk}^{(2)}$, and all of them are independent:

\begin{center}
   xzx, xxz, yyz, yzy, zxx, zyy, zzz (C2 axis along z direction)\\
\end{center}

\noindent For surface of the [100] domain (red), that is, the C2 axis is along [100] direction, one can write the susceptibility tensor as

\begin{equation}
\label{eq:S3}
\mathbf{\chi}_{ijk}^{(2)}(C_\mathrm{2v}) =
\begin{pmatrix}
\begin{bmatrix} \chi_{zzz} \\ 0 \\ 0 \end{bmatrix} &
\begin{bmatrix} 0 \\ \chi_{zyy} \\ 0 \end{bmatrix} &
\begin{bmatrix} 0 \\ 0 \\ \chi_{zxx} \end{bmatrix} \\[3.5em]

\begin{bmatrix} 0 \\ \chi_{yyz} \\ 0 \end{bmatrix} &
\begin{bmatrix} \chi_{yyz} \\ 0 \\ 0 \end{bmatrix} &
\begin{bmatrix} 0 \\ 0 \\ 0 \end{bmatrix} \\[3.5em]

\begin{bmatrix} 0 \\ 0 \\ \chi_{xxz} \end{bmatrix} &
\begin{bmatrix} 0 \\ 0 \\ 0 \end{bmatrix} &
\begin{bmatrix} \chi_{xxz} \\ 0 \\ 0 \end{bmatrix}
\end{pmatrix}
\end{equation}

\vspace{12pt}
\noindent For surface of the [010] domain (blue), that is, the C2 axis is along [010] direction, one can write the susceptibility tensor as

\begin{equation}
\label{eq:S5}
\mathbf{\chi}_{ijk}^{(2)}(C_\mathrm{2v}) =
\begin{pmatrix}
\begin{bmatrix} 0 \\ \chi_{xxz} \\ 0 \end{bmatrix} &
\begin{bmatrix} \chi_{xxz} \\ 0 \\ 0 \end{bmatrix} &
\begin{bmatrix} 0 \\ 0 \\ 0 \end{bmatrix} \\[3.5em]

\begin{bmatrix} \chi_{zxx} \\ 0 \\ 0 \end{bmatrix} &
\begin{bmatrix} 0 \\ \chi_{zzz} \\ 0 \end{bmatrix} &
\begin{bmatrix} 0 \\ 0 \\ \chi_{zyy} \end{bmatrix} \\[3.5em]

\begin{bmatrix} 0 \\ 0 \\ 0 \end{bmatrix} &
\begin{bmatrix} 0 \\ 0 \\ \chi_{yyz} \end{bmatrix} &
\begin{bmatrix} 0 \\ \chi_{yyz} \\ 0 \end{bmatrix}
\end{pmatrix}
\end{equation}

\vspace{12pt}
\noindent In normal incidence geometry, we simulate RA-SHG patterns in parallel and crossed polarization configurations.

\noindent For the surface of the [100] domain:

\begin{equation}
\label{eq:S4}
\begin{aligned}
I_{||}^{2\omega}(\varphi)[100] &= 
\left[ \chi_{zzz}\cos^3{\varphi} + (2\chi_{yyz} + \chi_{zyy})\cos{\varphi}\sin^2{\varphi} \right]^2 \\
I_{\perp}^{2\omega}(\varphi)[100] &= 
\left[ (2\chi_{yyz} - \chi_{zzz})\sin{\varphi}\cos^2{\varphi} - \chi_{zyy}\sin^3{\varphi} \right]^2
\end{aligned}
\end{equation}

\noindent For the surface of the [010] domain:

\begin{equation}
\label{eq:S6}
\begin{aligned}
I_{||}^{2\omega}(\varphi) [010]&= 
\left[ \chi_{zzz}\sin^3{\varphi} + (2\chi_{xxz} + \chi_{zxx})\sin{\varphi}\cos^2{\varphi} \right]^2 \\
I_{\perp}^{2\omega}(\varphi) [010]&= 
[ (2\chi_{xxz} - \chi_{zzz})\cos{\varphi}\sin^2{\varphi} -\chi_{zxx}\cos^3{\varphi}]^2
\end{aligned}
\end{equation}\\

The structural domain size is in the micrometer scale ~\cite{kalisky2013locally}, which is above the diffraction limit. Our beam size of $\sim$ 40 $\mu$m in diameter covers sufficient domains. As shown in \textbf{Fig.~\ref{SHG}(c)}, the RA-SHG pattern can be accounted in the superposition of ED-SHG intensity from three types of domain states with equal population.

\clearpage



\newpage
\section{$\textbf{S4. Magnetization characterization of LSMO/STO heterostructure}$}

\begin{figure*}[ht!]
\includegraphics[width=1.0\textwidth]{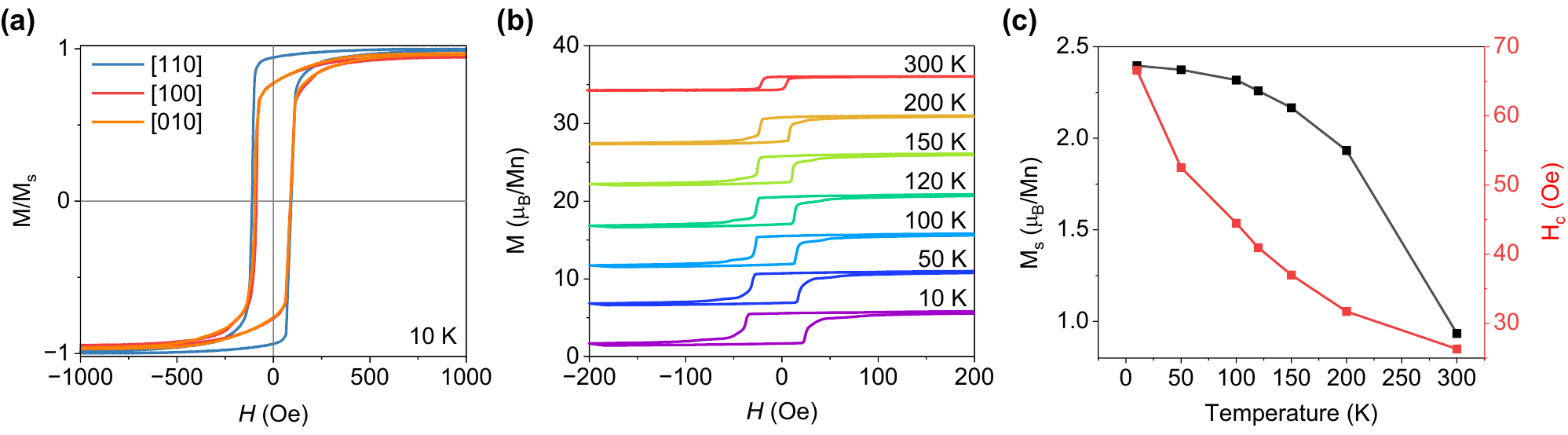}
\caption{\label{VSM} \justifying \small $\textbf{(a)}$ Magnetic hysteresis loops ($M-H$) with in-plane magnetic field applied along the [100] (red), [010] (orange), and [110] (blue) directions at 10 K. $\textbf{(b)}$ Magnetic hysteresis loops with in-plane magnetic field along [110] direction at varying temperatures. $\textbf{(c)}$ Saturation magnetization ($M_\mathrm{s}$) and coercive field ($H_\mathrm{c}$) as a function of temperature extracted from \textbf{(b)}.}
\end{figure*}

\textbf{Fig.~\ref{VSM}(a)} shows the magnetic hysteresis loops with in-plane magnetic field applied along the [100], [010], and [110] directions at 10 K. Our results confirm that the magnetic easy axis is along the [110] direction. \textbf{Fig.~\ref{VSM}(b)} shows magnetic hysteresis loops ranging from 10 K to 300 K. Saturation magnetization $\mathrm{M}_s$ and coercive field $\mathrm{H}_c$ extracted from \textbf{Fig.~\ref{VSM}(b)} are plotted as a function of temperature (see \textbf{Fig.~\ref{VSM}(c)}). The $M_\mathrm{s}$ = 2.4 $\mu_{\mathrm{B}}/\mathrm{Mn}$ and $H_\mathrm{c}$ = 66.5 Oe at 10 K are consistent with previously reported values in the 50 unit cell thickness LSMO thin films \cite{LSMOuB}. \\

\clearpage


\newpage
\section{$\textbf{S5. Methods of fitting FMR spectra}$ }

\begin{figure*}[ht!]
\includegraphics[width=0.95\textwidth]{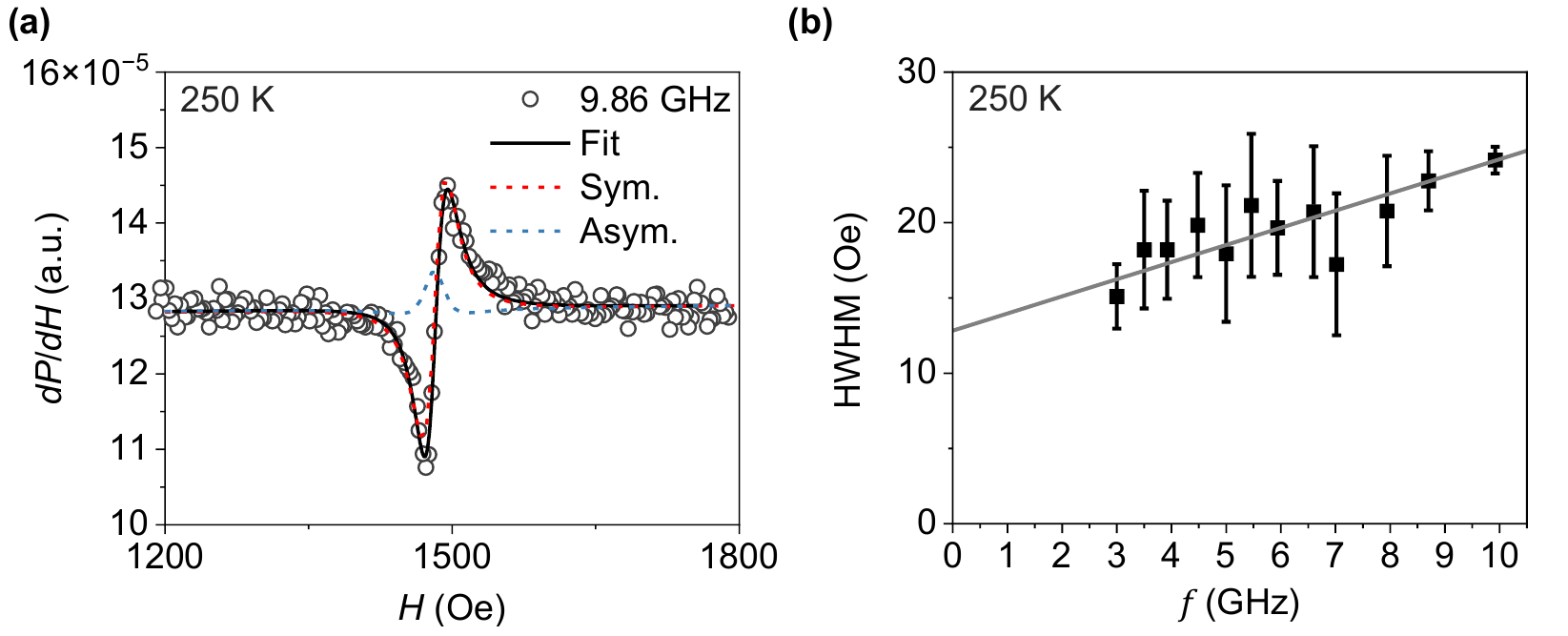}
\caption{\label{Damping} \justifying \small \textbf{(a)} FMR profile at 9.86 GHz acquired at 250 K. The fit to Eq.(S7) (black solid curve) and the decomposition of the symmetric (red dashed curve) and antisymmetric (blue dashed curve) components are overlaid on the raw data. \textbf{(b)} Line width as a function of frequency. Solid line is the fit to the model $\Delta H =\Delta H_0+\frac{2\pi}{\gamma}\alpha\mathit{f}$.}
\end{figure*}

The FMR line profile can be fitted to the derivative of a Lorentzian function, which can be decomposed into the superposition of symmetric and antisymmetric components \cite{sharma2021light}:

\begin{equation}
\label{eq:s7}
\frac{dP}{dH}= F_\mathrm{s}\frac{\Delta H(H-H_\mathrm{res})}{[(H-H_\mathrm{res})^{2}+\Delta H^{2}]^{2}}-F_\mathrm{a}\frac{{\Delta H}^{2}-(H-H_\mathrm{res})^{2}}{[(H-H_\mathrm{res})^{2}+{\Delta H}^{2}]^{2}}    
\end{equation}
where $H_\mathrm{res}$ is the resonance frequency, $\Delta H$ is the half-width at half-maximum (HWHM) linewidth. $F_\mathrm{s}$ and $F_\mathrm{a}$ are the intensity of symmetric and anti-symmetric components, respectively. \textbf{Fig.~\ref{Damping}(a)} shows the fit of the FMR line profile at 9.86 GHz as an example. As shown in \textbf{Fig.~\ref{Damping}(b)}, linewidth (HWHM) can be fitted to the model $\Delta H =\Delta H_0+\frac{2\pi}{\gamma}\alpha\mathit{f}$ \cite{kalarickal2006ferromagnetic}, where $\Delta H_0$ = 12.8 $\pm$ 1.9 Oe and $\alpha$ = 0.003 $\pm$ 0.001 are the inhomogeneous broadening and the effective damping constant, respectively. Therefore, in the magnon dissipation rate $\kappa_\mathrm{m}$ = $\Delta f_0+\alpha\mathit{f}$, the inhomogeneous broadening $\Delta f_0 = \frac{\gamma}{2\pi}\Delta H_0$ = 35.84 MHz at 250 K. \\

\newpage
\section{\textbf{S6. Splitting of the magnon bands at 80 K and 50 K}}

\begin{figure*}[ht!]
\includegraphics[width=1.0\textwidth]{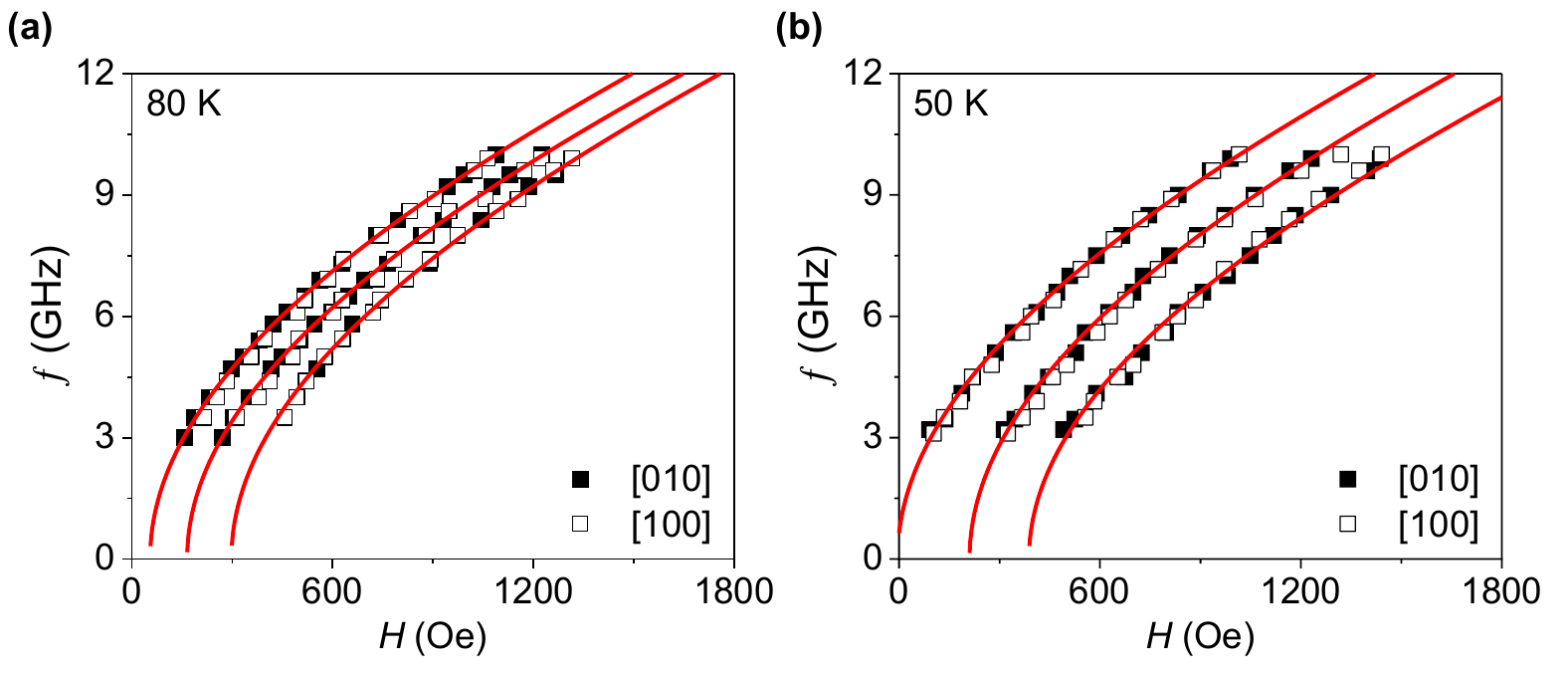}
\caption{\label{magnon} \justifying \small Split magnon bands of LSMO thin film at \textbf{(a)} 80 K and \textbf{(b)} 50 K, respectively. Red curves are fits to the Eq.(\ref{eq:s26a})-(\ref{eq:s26c}). Empty and solid squares are raw data from the measurements of $H\parallel[100]$ and $H\parallel[010]$, respectively.}

\end{figure*}

\textbf{Fig.~\ref{magnon}} shows the split magnon bands we extract from the raw FMR spectra. From 80 K to 50 K, the location of the middle band remains nearly unchanged, while the left and right magnon bands further separate apart. We also confirm that the results are consistent when the external magnetic field is applied along [100] and [010] directions.

Given that the thermal expansion coefficient of bulk \ce{SrTiO3} is 9.4$\times10^{-6}$ K$^{-1}$ ~\cite{lytle1964x}, the thermal-expansion-induced strain from 80 K to 50 K is 0.03\%. In comparison, the anisotropic local strain $\varepsilon_{c}-\varepsilon_{a}$ changes by 0.1\%, which plays the key role in the enhanced magnon band splitting from 80 K to 50 K.    

\newpage
\section{\textbf{S7. Analytical model of magnon-phonon coupling in LSMO/STO}}

We consider the LSMO/STO heterostructure as a 1D system where physical quantities only vary along the $z$ axis. The equation of motion that describes the mechanical displacement $u_{i} (z,t)$ can be written as

\begin{subequations}
\begin{equation}
\label{eq:s2a}
   \rho \frac{\partial ^{2}u_{x}(z,t)}{\partial t^{2}}=\frac{\partial }{\partial z}
(1+\beta \frac{\partial }{\partial t})\sigma _{xz}(z,t) 
\end{equation}

\begin{equation}
\label{eq:s2b}
   \rho \frac{\partial ^{2}u_{y}(z,t)}{\partial t^{2}}=\frac{\partial }{\partial z}
(1+\beta \frac{\partial }{\partial t})\sigma _{yz}(z,t) 
\end{equation}

\begin{equation}
\label{eq:s2c}
   \rho \frac{\partial ^{2}u_{z}(z,t)}{\partial t^{2}}=\frac{\partial }{\partial z}
(1+\beta \frac{\partial }{\partial t})\sigma _{zz}(z,t) 
\end{equation}
\end{subequations}

\vspace{5pt}
\noindent where $x\parallel[100]_{c}\equiv \text{\lq 1\rq}$, $y\parallel[010]_{c}\equiv \text{\lq 2\rq}$, and $z\parallel[001]_{c}\equiv \text{\lq 3\rq}$ represent the crystal coordinate system (subscription c: cubic); $\rho$ and $\beta$ are the mass density and the viscous elastic damping coefficient of the LSMO or STO. 

We consider the strained LSMO film in (pseudo)cubic structure and we calculate the stress tensor $\sigma _{iz}$ ($\textit{i}$ = $x$, $y$, $z$) using an elastic stiffness tensor $c_{ijkl}$ of cubic symmetry, that is, $\sigma_{ij}=c_{ijkl}(\varepsilon_{kl}-\varepsilon_{kl}^{0})$ \cite{PhysRevB.100.224405}, where the total strain can be expressed as $\varepsilon_{kl} = \frac{1}{2}(\frac{\partial{u_k}}{\partial l}+\frac{\partial{u_l}}{\partial_k})$ and the expression of the eigenstrain is given by $\varepsilon_{kl}^{0}=\frac{3}{2}\lambda_{111}(m_{k}m_{l})$~\cite{cullity2011introduction}, $k\neq{l}$ and $\varepsilon_{kl}^{0}=\frac{3}{2}\lambda_{100}(m_{k}^{2}-\frac{1}{3})$, $k=l$, with $\frac{3}{2}\lambda_{111}=-\frac{B_2}{2c_{44}}$ and $\frac{3}{2}\lambda_{100}=-\frac{B_1}{c_{11}-c_{12}}$. Here, the $\lambda_{111}$ and $\lambda_{100}$ are the magnetostrictive constants; $B_1$ and $B_2$ are the magnetoelastic coupling constants. 

The STO substrate undergoes a cubic-to-tetragonal structural phase transition as the sample is cooled below $T_\mathrm{S}$ = 105 K. Above $T_\mathrm{S}$, for the LSMO thin film, Eqs. (\ref{eq:s2a})-(\ref{eq:s2c}) can be rewritten as:

\begin{subequations}

\begin{equation}
\label{eq:s3a}
   \rho \frac{\partial ^{2}u_{x}(z,t)}{\partial t^{2}}-c_{44}\frac{\partial}{\partial {z^2}}(1+\beta\frac{\partial}{\partial t})u_{x}(z,t)=B_{2}\frac{\partial }{\partial z}
(1+\beta \frac{\partial }{\partial t})(m_{x}(z,t)m_{z}(z,t))
\end{equation}

\begin{equation}
\label{eq:s3b}
   \rho \frac{\partial ^{2}u_{y}(z,t)}{\partial t^{2}}-c_{44}\frac{\partial}{\partial {z^2}}(1+\beta\frac{\partial}{\partial t})u_{y}(z,t)=B_{2}\frac{\partial }{\partial z}
(1+\beta \frac{\partial }{\partial t})(m_{y}(z,t)m_{z}(z,t))
\end{equation}

\begin{equation}
\label{eq:s3c}
   \rho \frac{\partial ^{2}u_{z}(z,t)}{\partial t^{2}}-c_{11}\frac{\partial}{\partial {z^2}}(1+\beta\frac{\partial}{\partial t})u_{z}(z,t)=B_{1}\frac{\partial }{\partial z}
(1+\beta \frac{\partial }{\partial t})(m_{z}(z,t))
\end{equation}
\end{subequations}
Given that the magnon is Kittel mode, i.e. $\frac{\partial m_{x}(z,t)}{\partial z}=\frac{\partial m_{y}(z,t)}{\partial z}=\frac{\partial m_{z}(z,t)}{\partial z}=0$, the right sides of Eqs.(\ref{eq:s3a})-(\ref{eq:s3c}) are zero, which indicates that the magnetization oscillation $m_i (z,t)$ does not induce the acoustic phonon $u_i (z,t)$ within the LSMO film. In the nonmagnetic STO substrate, the equations of motion for $u_i (z,t)$ are similar to Eqs.(\ref{eq:s3a})-(\ref{eq:s3c}) except that the terms on the right sides are zero.

In the dynamical regime, we define mechanical displacement as $u_{i} (z,t) = u_{i}(t=0)+\Delta u_{i}(x,t)$ and magnetization oscillations in the form of plane wave perturbation $\textbf{m} =\textbf{m}^0+\Delta \textbf{m}(t)=(m_{x}^{0}+\Delta m_{x}^{0}e^{-\rm{i}\omega t}, \Delta m_{y}^{0}e^{-\rm{i}\omega t}, \Delta m_{z}^{0}e^{-\rm{i}\omega t})$, where $\textbf{m}^0=(m_{x}^{0},m_{y}^{0},m_{z}^{0})=(1,0,0)$ as the external static magnetic field $H$ is applied along the $x$ axis. Then, we can derive the stress distribution $\sigma_{iz}(z,t)=\sigma_{iz}(t=0)+\Delta \sigma_{iz}(z,t)$ in LSMO thin film (superscript ‘m’) and STO substrate (superscript ‘s’), respectively:

\begin{subequations}    
\begin{equation}
\label{eq:s4a}
   \Delta \sigma_{xz}^{m}(z,t)=c_{44}^{m} \partial\Delta u_{x}^{m}(z,t)/\partial z+B_{2}\Delta m_{z}^{0}e^{-i\omega t}; \Delta\sigma_{xz}^{s}(z)=c_{44}^{s}\partial \Delta u_{x}^{s}(z)/\partial z 
\end{equation}

\begin{equation}
\label{eq:s4b}
   \Delta \sigma_{yz}^{m}(z,t)=c_{44}^{m} \partial\Delta u_{y}^{m}(z,t)/\partial z; \Delta\sigma_{yz}^{s}(z)=c_{44}^{s}\partial \Delta u_{y}^{s}(z)/\partial z 
\end{equation}

\begin{equation}
\label{eq:s4c}
   \Delta \sigma_{zz}^{m}(z,t)=c_{11}^{m} \partial\Delta u_{z}^{m}(z,t)/\partial z; \Delta\sigma_{zz}^{s}(z)=c_{11}^{s}\partial \Delta u_{z}^{s}(z)/\partial z 
\end{equation}
\end{subequations}

Considering the stress-free condition (i.e., the stress continuity condition at the solid/air interface) at the top surface of the LSMO film ($z=-d$) and the bottom surface of the STO ($z=L$), i.e., $\Delta \sigma_{iz}^{m}(z=-d)=0$, and $\Delta \sigma_{iz}^{s}(z=L)=0$, and the stress/displacement continuity condition at the LSMO/STO interface, i.e. $\sigma_{iz}^{m}(z=0)=\sigma_{iz}^{s}(z=0)$, and $\Delta u_{i}^{m}(z=0)=\Delta u_{i}^{s}(z=0)$, with $i=x,y,z$, we can derive the analytical expressions of both the $\Delta u_{i}^{m}(z,t)$ and $\Delta u_{i}^{s}(z,t)$ that satisfy Eqs.(\ref{eq:s3a})-(\ref{eq:s3c}). Note that the Kittel mode magnon $\Delta m_{z}=\Delta m_{z}^{0}e^{-\rm{i}\omega t}$ is only associated with $\Delta \sigma_{xz}^{m}$ and $\Delta \sigma_{xz}^{s}$ (see Eq.(\ref{eq:s4a})), and thus can induce the $u_x (z,t)$ which varies spatially along $z$. This corresponds to the strain component $\epsilon_{xz}$ (TA phonon). In other words, the magnon can only generate TA phonons in the current setup at 250 K.

When the temperature decreases to 80 K, the STO substrate is tetragonal. In this case, the mechanical displacements of the LSMO thin film in the initial equilibrium state will be different from those at 250 K. For example, consider that  $x\parallel[100]_{c}\equiv \text{\lq 1\rq}$, $y\parallel[010]_{c}\equiv \text{\lq 2\rq}$, and $z\parallel[001]_{c}\equiv \text{\lq 3\rq}$ in the LSMO, and that $x\parallel[100]_{t}\equiv \text{\lq 1\rq}$, $y\parallel[010]_{t}\equiv \text{\lq 2\rq}$, and $z\parallel[001]_{t}\equiv \text{\lq 3\rq}$ (t: tetragonal), the initial condition of the mechanical displacements in the LSMO thin film becomes $u_x (t=0)=c_{STO}$ and $u_y (t=0)=a_{STO}$, where $a_{STO}$ and $c_{STO}$ are the lattice parameters of the STO at 80 K. For the pseudocubic LSMO film at 80 K, the equations of motion for the dynamic displacement $\Delta u_{i}(x,t)=u_{i} (x,t)-u_{i} (t=0)$ and the expressions of dynamical stress $\Delta \sigma_{iz} (z,t)$ keep the same with Eq.(S2) and Eq.(S3) at 250 K.

For the tetragonal STO with $x\parallel[100]_{t}\equiv \text{\lq 3\rq}$, the equations of motion and the dynamical stress can be written as,
\begin{subequations}
\begin{equation}
\label{eq:s5a}
   \rho \frac{\partial ^{2}u_{x}^{s}(z,t)}{\partial t^{2}}-c_{44}^{s}\frac{\partial}{\partial z^2}
(1+\beta ^{2} \frac{\partial }{\partial t})u_{x}^{s}(z,t)=0 
\end{equation}

\begin{equation}
\label{eq:s5b}
   \rho \frac{\partial ^{2}u_{y}^{s}(z,t)}{\partial t^{2}}-c_{66}^{s}\frac{\partial}{\partial z^2}
(1+\beta ^{2} \frac{\partial }{\partial t})u_{y}^{s}(z,t)=0 
\end{equation}

\begin{equation}
\label{eq:s5c}
   \rho \frac{\partial ^{2}u_{z}^{s}(z,t)}{\partial t^{2}}-c_{11}^{s}\frac{\partial}{\partial z^2}
(1+\beta ^{2} \frac{\partial }{\partial t})u_{z}^{s}(z,t)=0  
\end{equation}

\begin{equation}
\label{eq:s5d}
   \Delta \sigma _{xz}^{s}(z)= \frac{C_{44}^s\partial \Delta u_{x}^{s}(z)}{\partial z}, \Delta \sigma _{yz}^{s}(z)= \frac{C_{66}^s\partial \Delta u_{y}^{s}(z)}{\partial z}, \Delta \sigma _{zz}^{s}(z)= \frac{C_{11}^s\partial \Delta u_{z}^{s}(z)}{\partial z}
\end{equation}
\end{subequations}

For the tetragonal STO with $y\parallel[001]_{t}\equiv \text{\lq 3\rq}$, the equations of motion and the dynamical stress can be written as,

\begin{subequations}
\begin{equation}
\label{eq:s6a}
   \rho \frac{\partial ^{2}u_{x}^{s}(z,t)}{\partial t^{2}}-c_{66}^{s}\frac{\partial}{\partial z^2}
(1+\beta ^{2} \frac{\partial }{\partial t})u_{x}^{s}(z,t)=0 
\end{equation}

\begin{equation}
\label{eq:s6b}
   \rho \frac{\partial ^{2}u_{y}^{s}(z,t)}{\partial t^{2}}-c_{44}^{s}\frac{\partial}{\partial z^2}
(1+\beta ^{2} \frac{\partial }{\partial t})u_{y}^{s}(z,t)=0 
\end{equation}

\begin{equation}
\label{eq:s6c}
   \rho \frac{\partial ^{2}u_{z}^{s}(z,t)}{\partial t^{2}}-c_{11}^{s}\frac{\partial}{\partial z^2}
(1+\beta ^{2} \frac{\partial }{\partial t})u_{z}^{s}(z,t)=0  
\end{equation}

\begin{equation}
\label{eq:s6d}
   \Delta \sigma _{xz}^{s}(z)= \frac{C_{66}^s\partial \Delta u_{x}^{s}(z)}{\partial z}, \Delta \sigma _{yz}^{s}(z)= \frac{C_{44}^s\partial \Delta u_{y}^{s}(z)}{\partial z}, \Delta \sigma _{zz}^{s}(z)= \frac{C_{11}^s\partial \Delta u_{z}^{s}(z)}{\partial z}
\end{equation}

\end{subequations}

For the tetragonal STO with $z\parallel[001]_{t}\equiv \text{\lq 3\rq}$, the equations of motion and the dynamical stress can be written as,

\begin{subequations}
\begin{equation}
\label{eq:s7a}
   \rho \frac{\partial ^{2}u_{x}^{s}(z,t)}{\partial t^{2}}-c_{44}^{s}\frac{\partial}{\partial z^2}
(1+\beta ^{2} \frac{\partial }{\partial t})u_{x}^{s}(z,t)=0 
\end{equation}

\begin{equation}
\label{eq:s7b}
   \rho \frac{\partial ^{2}u_{y}^{s}(z,t)}{\partial t^{2}}-c_{44}^{s}\frac{\partial}{\partial z^2}
(1+\beta ^{2} \frac{\partial }{\partial t})u_{y}^{s}(z,t)=0 
\end{equation}

\begin{equation}
\label{eq:s7c}
   \rho \frac{\partial ^{2}u_{z}^{s}(z,t)}{\partial t^{2}}-c_{33}^{s}\frac{\partial}{\partial z^2}
(1+\beta ^{2} \frac{\partial }{\partial t})u_{z}^{s}(z,t)=0  
\end{equation}

\begin{equation}
\label{eq:s7d}
   \Delta \sigma _{xz}^{s}(z)= \frac{C_{44}^s\partial \Delta u_{x}^{s}(z)}{\partial z}, \Delta \sigma _{yz}^{s}(z)= \frac{C_{44}^s\partial \Delta u_{y}^{s}(z)}{\partial z}, \Delta \sigma _{zz}^{s}(z)= \frac{C_{33}^s\partial \Delta u_{z}^{s}(z)}{\partial z}
\end{equation}

\end{subequations}

\vspace{5pt}
Meanwhile, the stress/displacement continuity conditions at the surfaces and interfaces at 80 K remain the same as that at 250 K. As a result, the Kittel mode magnon $\Delta m_z= \Delta m_{z}^{0} e^{-\rm{i}\omega t}$ can likewise be only associated with the $\Delta \sigma _{xz}^{m}$ and $\Delta \sigma _{xz}^{s}$. This indicates that the magnon can only generate TA phonons in the current set-up at 80 K as well.


\newpage

\section{\textbf{S8. Analytical simulation of FMR spectra} $\textbf{P}_{\rm{abs}}(\omega, \textbf{H}^{\rm{bias}})$}

\noindent According to \cite{PhysRevB.104.014403,PhysRevLett.106.117601}, under the present condition, one has

\begin{equation}
\label{eq:s8}
   \textbf{P}_{\rm{abs}}(\omega , \textbf{H}^{\rm{bias}}) \propto \operatorname{Im}(\textbf{h}^{\ast \text{T}}\Delta\textbf{m})=\operatorname{Im}(\textbf{h}^{\ast \text{T}}\boldsymbol{\chi} \textbf{h}) = \operatorname{Im}(h_{i}^{0}\chi_{ij}h_{j}^{0})
\end{equation}

\noindent where the complex-valued magnetic susceptibility $\boldsymbol{\chi}$ is defined as, $\mathbf{\Delta}\textbf{m}=\boldsymbol{\chi} \textbf{h}$, with $h_{i}=h_{i}^{0} e^{-\rm{i}\omega t} (i=x,y,z)$ being spatially uniform in the magnet and its conjugate $h_{i}^{\ast}=h_{i}^{0} e^{-\rm{i}\omega t}$ . For the present set-up, Eq.(\ref{eq:s8}) reduces to $\textbf{P}_{\rm{abs}} \propto \operatorname{Im}(h_{y}^{{0}^{2}} \chi_{yy} ) \propto \operatorname{Im}({\chi}_{yy})$.  The susceptibility tensor $\boldsymbol{\chi}$ can be computed analytically by solving the linearized LLG equation and the elastodynamic equation. The LLG equation is written as
\begin{equation}
\label{eq:s9}
   \frac{\partial\textbf{m}}{\partial t} = -\gamma(\textbf{m}\times(\textbf{H}^{\rm{eff}}+\textbf{h}))+\alpha(\textbf{m}\times\frac{\partial\textbf{m}}{\partial t})
\end{equation}

\noindent where the effective field $\textbf{H}^{\rm{eff}}=\textbf{H}^{\rm{anis}}+\textbf{H}^{\rm{d}}+\textbf{H}^{\rm{mel}}+\textbf{H}^{\rm{bias}}+\textbf{h}$, where $\textbf{H}^{\rm{d}} = (0,0,-M_{s}m_{z}$ is the demagnetization field,  $\textbf{H}^{\rm{anis}}$ is the magnetocystalline anisotropy effective field, and $\textbf{H}^{\rm{mel}}$ is the magnetoelastic effective field. For studying Kittel mode magnon, the exchange coupling field $\textbf{H}^{\rm{exch}}$ does not to be $\textbf{H}^{\rm{anis}}$ is given by
\begin{equation}
\label{eq:s10}
   \text{H}_{i}^{\rm{anis}}=-\frac{2m_{i}}{\mu _{0}M_{s}}[K_{1}(m_{j}^{2}+m_{k}^{2})], i,j,k=x,y,z
\end{equation}

\noindent where the $i,j,k$ are different from each other and the high-order anisotropy coefficients ($K_2$ and above) are omitted (as in the case in (001) LSMO \cite{PhysRevB.83.134408}). $\textbf{H}^{\text{mel}}$ is the magnetoelastic effective field, given as
\begin{equation}
\label{eq:s11}
    \text{H}_{i}^{\rm{mel}}=-\frac{2}{\mu _{0}M_{s}}[B_{1}m_{i}\varepsilon_{ii}+B_{2}(m_{j}\varepsilon _{ij}+m_{k}\varepsilon _{ik})]
\end{equation}

\noindent where $\varepsilon_{ij}$ is the total strain, describing the deformation with respect to the higher-symmetry paramagnetic phase. For dynamical excitation, one has $\varepsilon_{ij}= \varepsilon_{ij} (t=0)+\Delta\varepsilon_{ij} (z,t)$, where $\varepsilon_{ij} (t=0)$ is the total strain in the LSMO in the initial equilibrium state and can be calculated based on the mechanical boundary condition. Specifically, for a coherently strained $(001)_{\rm{pc}}$ LSMO thin film, one has \cite{PhysRevB.80.224416}, 
\begin{subequations}
\begin{equation}
\label{eq:s12a}
   \varepsilon_{xx}(t=0)=\varepsilon_{xx}^{\rm{mis}},\varepsilon_{yy}(t=0)=\varepsilon_{yy}^{\rm{mis}},\varepsilon_{xy}(t=0)=0
\end{equation}

\begin{equation}
\label{eq:s12b}
   \varepsilon_{zz}(t=0)=\frac{B_{1}-3c_{12}(\varepsilon_{xx}^{\rm{mis}}+\varepsilon_{yy}^{\rm{mis}})}{3c_{11}},\varepsilon_{yz}(t=0)=0,\varepsilon_{xz}(t=0)=0
\end{equation}

\end{subequations}

\noindent where $m_i^0 (i=x,y,z)$ refers to the magnetization in the initial equilibrium state. For the present 1D system where physical quantities only vary along the $z$ axis, the dynamical strain $\Delta\varepsilon_{ij} (z,t)$ can be expanded as

\begin{subequations}
\begin{equation}
\label{eq:s13a}
   \Delta \varepsilon_{xx}(z,t)=0,\Delta \varepsilon_{yy}(z,t)=0,\Delta \varepsilon_{yy}(z,t)=\frac{\partial \Delta u_{z}(z,t)}{\partial z} 
\end{equation}

\begin{equation}
\label{eq:s13b}
   \Delta \varepsilon_{yz}(z,t)=\frac{1}{2} \frac{\partial \Delta u_{y}(z,t)}{\partial z},\Delta \varepsilon_{xz}(z,t)=\frac{1}{2} \frac{\partial \Delta u_{x}(z,t)}{\partial z},\Delta \varepsilon_{xy}(z,t)=0
\end{equation}

\end{subequations}

\noindent By writing the variation of the $\textbf{m}$ around the initial equilibrium state $\textbf{m}^0$ in the form of a plane-wave perturbation, i.e., $\textbf{m}=\textbf{m}^0+\Delta\textbf{m}(t)=(m_{x}^{0}+\Delta m_{x}^{0} e^{-\rm{i}\omega t},\Delta m_{y}^{0} e^{-\rm{i}\omega t},\Delta m_{z}^{0} e^{-\rm{i}\omega t})$, substituting $\textbf{m}$ into Eq.(\ref{eq:s9}), and dropping the higher-order terms, the latter can be rewritten into a linearized form,

\begin{equation}
\label{eq:s14}
    \begin{pmatrix}
 \rm{i}\omega & 0  & 0  \\
 0 & \rm{i}\omega  & A_{23}  \\
 0 & A_{32} & \rm{i}\omega  \\
\end{pmatrix}
\begin{pmatrix}
 \Delta m_{x}^{0}e^{-\rm{i}\omega t}\\
 \Delta m_{y}^{0}e^{-\rm{i}\omega t}\\
 \Delta m_{z}^{0}e^{-\rm{i}\omega t}
\end{pmatrix} = \gamma
\begin{pmatrix}
 0\\
 0\\
 h_{y}^{0}e^{-\rm{i}\omega t}
\end{pmatrix} +
\begin{pmatrix}
 0\\
 \Omega_{y}(z,t)\\
0 
\end{pmatrix}
\end{equation}
where,

\begin{subequations}

\begin{eqnarray}
\begin{aligned}
\label{eq:s15a}
{A_{23}}& =  {\rm{i}}\alpha \omega  - \gamma H_x^{{\rm{bias}}} - \gamma {M_s}\\ & - \frac{{2\gamma {K_1}}}{{{\mu _0}{M_s}}} - \frac{{2\gamma B_1^2}}{{3{c_{11}}{\mu _0}{M_s}}} + \frac{{2\gamma B_2^2}}{{{c_{44}}{\mu _0}{M_s}}} \\ &+ \frac{{2\gamma {B_1}\varepsilon _{xx}^{{\rm{mis}}}}}{{{\mu _0}{M_s}}} + \frac{{2\gamma {B_1}{c_{12}}\varepsilon _{xx}^{{\rm{mis}}}}}{{{c_{11}}{\mu _0}{M_s}}} + \frac{{2\gamma {B_1}{c_{12}}\varepsilon _{yy}^{{\rm{mis}}}}}{{{c_{11}}{\mu _0}{M_s}}}
\end{aligned}
\end{eqnarray}

\begin{equation}
\label{eq:s15b}
{A_{32}} =  - {\rm{i}}\alpha \omega  + \gamma H_x^{{\rm{bias}}} + \frac{{2\gamma {K_1}}}{{{\mu _0}{M_s}}} - \frac{{2\gamma {B_1}\varepsilon _{xx}^{{\rm{mis}}}}}{{{\mu _0}{M_s}}} + \frac{{2\gamma {B_1}\varepsilon _{yy}^{{\rm{mis}}}}}{{{\mu _0}{M_s}}}
\end{equation}

\begin{equation}
\label{eq:s15c}
{{\rm{\Omega }}_y}\left( {z,t} \right) = \frac{{\gamma {B_2}}}{{{\mu _0}{M_s}}}\frac{{\partial \Delta {u_x}\left( {z,t} \right)}}{{\partial z}} = \frac{{2\gamma {B_2}}}{{{\mu _0}{M_s}}}\Delta {\varepsilon _{xz}}\left( {z,t} \right)
\end{equation}
\end{subequations}

\noindent where the terms $\Omega_y (z,t)$ characterizes the bidirectional magnon-phonon coupling (unit: Hz) because the TA phonon, represented by $\Delta\varepsilon_{xz} (z,t)$, is generated by the magnons and in turn modulates the magnetization dynamics. 

\noindent From the above writing, it has been demonstrated that the harmonic component of the $\Delta u_x (z,t)$ can be expressed as the function of the magnetization $\Delta m_z (t)=\Delta m_{z}^0 e^{-\rm{i}\omega t}$, which means that we can rewrite the coupling strength $\Omega_y (z,t)$ as $\Omega_y (z,t)=\vartheta_{yz} \Delta m_{z} (t)=\vartheta_{yz} \Delta m_{z}^{0} e^{-\rm{i}\omega t}$ by introducing an auxiliary coefficient matrix $\vartheta_{yj}$. Substituting this expression of $\Omega_y (z,t)$ into Eq.(\ref{eq:s14}), Eq.(\ref{eq:s14}) can be rewritten into,
\begin{subequations}
\begin{equation}
\label{eq:s16a}
\left( {\begin{array}{*{20}{c}}
{\Delta m_x^0{e^{ - {\rm{i}}\omega t}}}\\
{\Delta m_y^0{e^{ - {\rm{i}}\omega t}}}\\
{\Delta m_z^0{e^{ - {\rm{i}}\omega t}}}
\end{array}} \right) = \boldsymbol{\chi} \left( {\begin{array}{*{20}{c}}
0\\
{h_y^0{e^{ - {\rm{i}}\omega t}}}\\
0
\end{array}} \right)
\end{equation}

\begin{equation}
\label{eq:s16b}
{\boldsymbol{\chi }} = \gamma {\left( {\begin{array}{*{20}{c}}
{{\rm{i}}\omega }&0&0\\
0&{{\rm{i}}\omega }&{{A_{23}} - {\vartheta _{yz}}}\\
0&{{A_{32}}}&{{\rm{i}}\omega }
\end{array}} \right)^{ - 1}}\left( {\begin{array}{*{20}{c}}
0&0&0\\
0&0&{ - 1}\\
0&1&0
\end{array}} \right)
\end{equation}
\end{subequations}
\noindent The only unknown parameters in the expression of $\boldsymbol{\chi}$ are $\vartheta_{yi} (i=x,y,z)$, which can be calculated based on ${{\rm{\Omega }}_y}\left( {z,t} \right) = \frac{{2\gamma {B_2}}}{{{\mu _0}{M_s}}}\Delta {\varepsilon _{xz}}\left( {z,t} \right) = {\vartheta _{yz}}\left( z \right)\Delta m_z^0{e^{ - {\rm{i}}\omega t}}$, if the analytical expression of $\Delta \varepsilon_{xz} (z,t)$ is known. The $\Delta \varepsilon_{xz} (z,t)$  can be obtained by analytically solving the elastodynamic equation for $u_{x} (z,t)$ in the 1D system under the stress and displacement continuity conditions. 

\noindent As discussed earlier, the magnons can only generate TA phonons $u_x (z,t)$ in the present LSMO/STO system. Consider that $u_x (z,t)=u_x (t=0)+\Delta u_x (x,t)$, the equation of motion for $u_x (z,t)$, Eq.(\ref{eq:s3a}), at 250 K, can be rewritten as

\begin{equation}
\label{eq:s17}
   \rho \frac{{{\partial ^2}\Delta {u_x}\left( {z,t} \right)}}{{\partial {t^2}}} - {c_{44}}\frac{\partial }{{\partial {{\rm{z}}^2}}}\left( {1 + \beta \frac{\partial }{{\partial t}}} \right)\Delta {u_x}\left( {z,t} \right) = 0
\end{equation}

\noindent where the coefficients vary with the location $z$, i.e., either for the LSMO or the STO. For the present 1D system with plane-wave assumption, the $\Delta u_x$ that satisfies Eq.(\ref{eq:s17}) take the form of
\begin{subequations}

\begin{equation}
\label{eq:s18a}
   \Delta u_x^{\rm{m}}\left( {z,t} \right) = \Delta u_x^{{\rm{m}} + }{e^{\rm{i}\left( {k_{\rm{T}}^{\rm{m}}z - \omega t} \right)}} + \Delta u_x^{{\rm{m}} - }{e^{ - \rm{i}\left( {k_{\rm{T}}^{\rm{m}}z + \omega t} \right)}}
\end{equation}

\begin{equation}
\label{eq:s18b}
\Delta u_x^{\rm{s}}\left( {z,t} \right) = \Delta u_x^{{\rm{s}} + }{e^{\rm{i}\left( {k_{\rm{T}}^{\rm{s}}z - \omega t} \right)}} + \Delta u_x^{{\rm{s}} - }{e^{ - \rm{i}\left( {k_{\rm{T}}^{\rm{s}}z + \omega t} \right)}}
\end{equation}
\end{subequations}

\noindent where the $\Delta u_x^{{\rm{m}} \pm }$ and $\Delta u_x^{{\rm{s}} \pm }$) are the amplitudes of the forward-propagating (wavevector $\textbf{k}\parallel + z$) and backward propagating ($\textbf{k}\parallel - z$) acoustic waves in the LSMO and STO substrate, respectively. The transverse wavenumber in the LSMO magnet can be written as, 
$k_{\rm{T}}^{\rm{m}} = \sqrt {\frac{{{\rho ^{\rm{m}}}}}{{c_{44}^{\rm{m}}}}\frac{{{\omega ^2}}}{{1 - {\rm{i}}{\beta ^{\rm{m}}}\omega }}}  \approx \sqrt {\frac{{{\rho ^{\rm{m}}}}}{{c_{44}^{\rm{m}}}}} \omega \left( {1 + \frac{{\rm{i}}}{2}{\beta ^{\rm{m}}}\omega } \right)$, and that in the STO substrate can be written as $k_{\rm{T}}^{\rm{s}} = \sqrt {\frac{{{\rho ^{\rm{s}}}}}{{c_{44}^{\rm{s}}}}\frac{{{\omega ^2}}}{{1 - i{\beta ^{\rm{s}}}\omega }}}  \approx \sqrt {\frac{{{\rho ^{\rm{s}}}}}{{c_{44}^{\rm{s}}}}} \omega \left( {1 + \frac{{\rm{i}}}{2}{\beta ^{\rm{s}}}\omega } \right)$.
And we can derive the stress distribution in the magnet and the substrate:
\begin{subequations}

\begin{equation}
\label{eq:s19a}
   \sigma _{xz}^{\rm{m}}\left( {z,t} \right) = c_{44}^{\rm{m}}\partial \Delta u_x^{\rm{m}}\left( {z,t} \right)/\partial z + {B_2}\Delta {m_z}\left( t \right)
\end{equation}

\begin{equation}
\label{eq:s19b}
   \sigma _{xz}^{\rm{s}}\left( z \right) = c_{44}^{\rm{s}}\partial \Delta u_x^{\rm{s}}\left( z \right)/\partial z
\end{equation}
\end{subequations}

\noindent Considering the stress-free boundary condition: $\sigma_{xz}^m (z=-d)=0$ , and $\sigma_{xz}^s (z=L)=0$, and the continuous boundary condition in the interface, i.e. $\sigma_{xz}^m (z=0)=\sigma_{xz}^s (z=0)$, and $\Delta u_{x}^{m} (z=0)=\Delta u_{x}^{s} (z=0)$, with $i=x,y,z$, we can derive the analytical expressions of both the $\Delta u_{x}^{m \pm}$ and $\Delta u_{x}^{s \pm}$, both of which are function of the magnetization $\Delta m_{z} (t)=\Delta m_{z}^{0} e^{-\rm{i}\omega t}$

\noindent A knowledge of $\Delta u_{x}^{m \pm}$ allows us to calculate $\Delta \varepsilon _{xz} (z,t)$ via Eq.(\ref{eq:s19a}). The explicit expression of $\Delta \varepsilon_{xz} (z,t)$ in the LSMO film can be expressed as
\begin{eqnarray}
\begin{aligned}
&\Delta {\varepsilon _{xz}}\left( {z,t} \right)  = \\&
   - \frac{{{B_2}}}{{2c_{44}^{\rm{m}}}}\frac{{2c_{44}^{\rm{m}}k_{\rm{T}}^{\rm{m}}\cos \left( {k_{\rm{T}}^{\rm{s}}L} \right)\cos \left( {\frac{1}{2}k_{\rm{T}}^{\rm{m}}\left( {d + 2z} \right)} \right)\sin \left( {\frac{1}{2}k_{\rm{T}}^{\rm{m}}d} \right) + c_{44}^{\rm{s}}k_{\rm{T}}^{\rm{s}}\cos \left( {k_{\rm{T}}^{\rm{m}}z} \right)\sin \left( {k_{\rm{T}}^{\rm{s}}L} \right)}}{{c_{44}^{\rm{m}}k_{\rm{T}}^{\rm{m}}\cos \left( {k_{\rm{T}}^{\rm{s}}L} \right)\sin \left( {k_{\rm{T}}^{\rm{m}}d} \right) + c_{44}^{\rm{s}}k_{\rm{T}}^{\rm{s}}\cos \left( {k_{\rm{T}}^{\rm{m}}d} \right)\sin \left( {k_{\rm{T}}^{\rm{s}}L} \right)}}\\&\times\Delta m_z^0{e^{ - {\rm{i}}\omega t}},\ \ \ \ \ \ \ \ \ \ \ \     - d < z < 0\
\label{eq:s20}
\end{aligned}
\end{eqnarray}

\noindent which varies along the $z$ axis. The coefficient $\vartheta _{yz}$ in the LSMO can be calculated by
\begin{eqnarray}
\label{eq:s21}
\begin{aligned}
    \vartheta _{yz}\left( z \right) &= \frac{{\frac{{2\gamma {B_2}}}{{{\mu _0}{M_s}}}\Delta {\varepsilon _{xz}}\left( {z,t} \right)}}{\Delta m_{z}^{0} e^{ - \rm{i}\omega t}} \\ &=  - \frac{{\gamma B_2^2}}{{{\mu _0}{M_s}c_{44}^{\rm{m}}}} \\& \times \frac{{2c_{44}^{\rm{m}}k_{\rm{T}}^{\rm{m}}\cos \left( {k_{\rm{T}}^{\rm{s}}L} \right)\cos \left( {\frac{1}{2}k_{\rm{T}}^{\rm{m}}\left( {d + 2z} \right)} \right)\sin \left( {\frac{1}{2}k_{\rm{T}}^{\rm{m}}d} \right) + c_{44}^{\rm{s}}k_{\rm{T}}^{\rm{s}}\cos \left( {k_{\rm{T}}^{\rm{m}}z} \right)\sin \left( {k_{\rm{T}}^{\rm{s}}L} \right)}}{{c_{44}^{\rm{m}}k_{\rm{T}}^{\rm{m}}\cos \left( {k_{\rm{T}}^{\rm{s}}L} \right)\sin \left( {k_{\rm{T}}^{\rm{m}}d} \right) + c_{44}^{\rm{s}}k_{\rm{T}}^{\rm{s}}\cos \left( {k_{\rm{T}}^{\rm{m}}d} \right)\sin \left( {k_{\rm{T}}^{\rm{s}}L} \right)}},
    \\ & - d < z < 0
\end{aligned}
\end{eqnarray}

\noindent If assuming that the acoustic wave is largely uniform in the LSMO, we can further write,
\begin{subequations}
\begin{eqnarray}
\label{eq:s22a}
\begin{aligned}
    \Delta {\varepsilon _{xz}}\left( t \right) &\approx \Delta \varepsilon _{xz}^{\rm{m}}\left( z \right){e^{ - {\rm{i}}\omega t}} \\&= \frac{{{\mu _0}{M_s}}}{{2\gamma {B_2}}}{\vartheta _{yz}}\left( z \right)\Delta m_z^0{e^{ - {\rm{i}}\omega t}} \\&= \frac{{{\mu _0}{M_s}}}{{2\gamma {B_2}d}}\left( {\mathop \smallint \nolimits_{ - d}^0 {\vartheta _{yz}}\left( z \right)dz} \right)\Delta m_z^0{e^{ - {\rm{i}}\omega t}}
\end{aligned}
\end{eqnarray}

\begin{eqnarray}
\label{eq:s22b}
\begin{aligned}
    {\vartheta _{yz}}\left( z \right) =  - \frac{{\gamma B_2^2}}{{{\mu _0}{M_s}c_{44}^{\rm{m}}k_{\rm{T}}^{\rm{m}}d}}\frac{{4c_{44}^{\rm{m}}k_{\rm{T}}^{\rm{m}}\cos \left( {k_{\rm{T}}^{\rm{s}}L} \right){{\sin }^2}\left( {k_{\rm{T}}^{\rm{m}}d/2} \right) + c_{44}^{\rm{s}}k_{\rm{T}}^{\rm{s}}\sin \left( {k_{\rm{T}}^{\rm{m}}d} \right)\sin \left( {k_{\rm{T}}^{\rm{s}}L} \right)}}{{c_{44}^{\rm{m}}k_{\rm{T}}^{\rm{m}}\cos \left( {k_{\rm{T}}^{\rm{s}}L} \right)\sin \left( {k_{\rm{T}}^{\rm{m}}d} \right) + c_{44}^{\rm{s}}k_{\rm{T}}^{\rm{s}}\cos \left( {k_{\rm{T}}^{\rm{m}}d} \right)\sin \left( {k_{\rm{T}}^{\rm{s}}L} \right)}}
\end{aligned}
\end{eqnarray}

\noindent using the value of $\left\langle {{\vartheta _{yz}}\left( z \right)} \right\rangle $ to substitute the $\vartheta _{yz}$ in Eq.(\ref{eq:s16b}), one can calculate the full susceptibility tensor $\boldsymbol{\chi}$, and therefore the microwave absorption power via Eq.(\ref{eq:s8}).

\noindent At 80 K, for a tetragonal STO with $x\parallel[001]_{t}\equiv \text{\lq 3\rq}$ or $z\parallel[001]_{t}\equiv \text{\lq 3\rq}$, the equation of motion for the $u_{x}^{s} (z,t)$ and expression of the dynamical stress $\Delta \sigma_{xz}^{s} (z)$ keep the same as the Eq.(\ref{eq:s17}) and Eq.(\ref{eq:s19b}). The final expression of the full susceptibility tensor remains the same with that at 250K. However, for a tetragonal STO with $y\parallel[001]_{t}\equiv \text{\lq 3\rq}$, the $c_{44}$ in Eq.(\ref{eq:s17}) should be replaced by the $c_{66}$ of the STO while the $c_{44}^{S}$ in Eq.(\ref{eq:s19b}) should be replaced by $c_{66}^{S}$. Consequently, the expression of \[\left\langle {{\vartheta _{yz}}\left( z \right)} \right\rangle \] needs to be updated to,
\begin{eqnarray}
\label{eq:s22c}
\begin{aligned}
    {\vartheta _{yz}}\left( z \right) =  - \frac{{\gamma B_2^2}}{{{\mu _0}{M_s}c_{44}^{\rm{m}}k_{\rm{T}}^{\rm{m}}d}}\frac{{4c_{44}^{\rm{m}}k_{\rm{T}}^{\rm{m}}\cos \left( {k_{\rm{T}}^{\rm{s}}L} \right){{\sin }^2}\left( {k_{\rm{T}}^{\rm{m}}d/2} \right) + c_{66}^{\rm{s}}k_{\rm{T}}^{\rm{s}}\sin \left( {k_{\rm{T}}^{\rm{m}}d} \right)\sin \left( {k_{\rm{T}}^{\rm{s}}L} \right)}}{{c_{44}^{\rm{m}}k_{\rm{T}}^{\rm{m}}\cos \left( {k_{\rm{T}}^{\rm{s}}L} \right)\sin \left( {k_{\rm{T}}^{\rm{m}}d} \right) + c_{66}^{\rm{s}}k_{\rm{T}}^{\rm{s}}\cos \left( {k_{\rm{T}}^{\rm{m}}d} \right)\sin \left( {k_{\rm{T}}^{\rm{s}}L} \right)}}
\end{aligned}
\end{eqnarray}
\end{subequations}
\noindent where $k_{\rm{T}}^{\rm{s}} = \sqrt {\frac{{{\rho ^{\rm{s}}}}}{{c_{66}^{\rm{s}}}}\frac{{{\omega ^2}}}{{1 - i{\beta ^{\rm{s}}}\omega }}}  \approx \sqrt {\frac{{{\rho ^{\rm{s}}}}}{{c_{66}^{\rm{s}}}}} \omega \left( {1 + \frac{{\rm{i}}}{2}{\beta ^{\rm{s}}}\omega } \right)$.


\newpage
\section{$\textbf{S9. Analytical simulation of magnon band at 250 K and 80 K}$ }

Neglecting the coupling between dynamic strains and spin waves, i.e. $\Omega_{y} (z,t)=0$, the absorption spectrum $\textbf{P}_{\rm{abs}}$ will exhibit the characteristic figure of an FMR experiment. Under each bias magnetic field, the strongest absorption, i.e., the maximum value of ${\rm{Im}}\left( {v} \right)$, happens at the frequency $\omega_{\rm{m}}$. From Eq.(\ref{eq:s16b}), one can evaluate that ${\boldsymbol{\chi }} = {\chi _{yy}} = \frac{{\gamma {A_{23}}}}{{{A_{23}}{A_{32}} + {\omega ^2}}}$, which maximizes when the real component of the denominator is zero, i.e., ${\rm{Re}}\left( {{A_{23}}{A_{32}} + \omega _{\rm{m}}^2} \right) = 0$. Accordingly, the expression of the FMR frequency $\omega_{\rm{m}}$ can be written as,
\begin{equation}
\label{eq:s23}
{\omega _m} = \sqrt { - {\rm{Re}}\left( {{A_{23}}{A_{32}}} \right)} 
\end{equation}
                                        
Plugging in the expression of the $A_{23}$ and $A_{32}$ (Eqs.(\ref{eq:s15a})-(\ref{eq:s15b})) into Eq.(\ref{eq:s23}), one has

\begin{eqnarray}
\label{eq:s24}
    \begin{aligned}
        {\omega _{\rm{m}}} &= \frac{\gamma }{{{\mu _0}{M_s}}}\sqrt {\frac{1}{{3{c_{11}}{c_{44}}}}}  \\& \sqrt {\left( {2{B_1}\left( {{\varepsilon _{yy}} - {\varepsilon _{xx}}} \right) + {\mu _0}{M_s}H_x^{{\rm{bias}}} + 2{K_1}} \right)\left( { - 6{B_1}{c_{44}}\left( {{c_{11}} + {c_{12}}} \right){\varepsilon _{xx}} - 6{B_1}{c_{12}}{c_{44}}{\varepsilon _{yy}} + {A_0}} \right)} 
    \end{aligned}
\end{eqnarray}
\noindent where ${A_0} =  - 3{c_{11}}B_2^2 + 2{c_{44}}B_1^2 + 3{c_{11}}{c_{44}}\left( {2{K_1} + {\mu _0}{M_s}\left( {H_x^{{\rm{bias}}} + {M_s}} \right)} \right)$.\\

\noindent\textbf{(I)} For STO in the cubic phase at 250 K, we have ${\varepsilon _{xx}} = {\varepsilon _{yy}} = {\varepsilon ^{{\rm{mis}}}} = \frac{{{a_{{\rm{STO}}}} - {a_{{\rm{LSMO}}}}}}{{{a_{{\rm{LSMO}}}}}}$, the formula reduces to
\begin{eqnarray}
    \label{eq:s25}
    \begin{aligned}
        \omega _{\rm{m}}^{\rm{cubic}} = \frac{\gamma }{{{\mu _0}{M_s}}}\sqrt {\frac{{{\mu _0}{M_s}H_x^{{\rm{bias}}} + 2{K_1}}}{{3{c_{11}}{c_{44}}}}} \sqrt { - 6{B_1}{c_{44}}\left( {{c_{11}} + 2{c_{12}}} \right){\varepsilon ^{{\rm{mis}}}} + {A_0}} 
    \end{aligned}
\end{eqnarray}

Note that, at 250K, the value of the parameter $K_{1}$ remains unknown, we extracted 10 data points from the experimental FMR curve and defined the loss function as the sum of the squared differences between the calculated and experimentally measured FMR frequencies, i.e., $\sum {\left( {{\omega _{{\rm{m}},{\rm{cal}}}} - {\omega _{{\rm{m}},{\rm{exp}}}}} \right)^2}$. We applied the limited-memory BFGS algorithm to optimize the parameter $K_{1}$ via the Python’s SciPy package. Starting from an initial guess of $-$5000 $\textrm{J/m}^3$, the algorithm iteratively minimizes the loss function until convergence, ultimately returning the optimized value of $K_{1}=-$1766 $\textrm{J/m}^3$ (see Table S1).\\

\noindent\textbf{(II)} For STO in the tetragonal phase at 80 K, the interfacial strain becomes anisotropic. Thus, we consider the following three cases:

\noindent\textbf{(a)} When $x\parallel [001]_{t}$ in the STO substrate ($x\parallel[100]_{c}$), we have ${\varepsilon _{xx}} = \frac{{{a_{{\rm{STO}}}} - {a_{{\rm{LSMO}}}}}}{{{a_{{\rm{LSMO}}}}}} = {\varepsilon _a}$, ${\varepsilon _{yy}} = \frac{{{c_{{\rm{STO}}}} - {a_{{\rm{LSMO}}}}}}{{{a_{{\rm{LSMO}}}}}} = {\varepsilon _c}$, and, 
\begin{subequations}
\begin{eqnarray}
    \label{eq:s26a}
    \begin{aligned}
        \omega _{\rm{m}}^{\left[ {100} \right]} &= \frac{\gamma }{{{\mu _0}{M_s}}}\sqrt {\frac{1}{{3{c_{11}}{c_{44}}}}} \\& \sqrt {\left( {2{B_1}\left( {{\varepsilon _c} - {\varepsilon _a}} \right) + {\mu _0}{M_s}H_x^{{\rm{bias}}} + 2{K_1}} \right)\left( { - 6{B_1}{c_{44}}\left( {{c_{11}} + {c_{12}}} \right){\varepsilon _a} - 6{B_1}{c_{12}}{c_{44}}{\varepsilon _c} + {A_0}} \right)} 
    \end{aligned}
\end{eqnarray}

\noindent\textbf{(b)} When $y\parallel[001]_{t}$ in the STO substrate ($y\parallel[010]_{c}$), we have $\varepsilon_{xx}=\varepsilon_{c}$, $\varepsilon_{yy}=\varepsilon_{a}$, and,
\begin{eqnarray}
    \label{eq:s26b}
    \begin{aligned}
        \omega _{\rm{m}}^{\left[ {010} \right]} &= \frac{\gamma }{{{\mu _0}{M_s}}}\sqrt {\frac{1}{{3{c_{11}}{c_{44}}}}} \\& \sqrt {\left( {2{B_1}\left( {{\varepsilon _a} - {\varepsilon _c}} \right) + {\mu _0}{M_s}H_x^{{\rm{bias}}} + 2{K_1}} \right)\left( { - 6{B_1}{c_{44}}\left( {{c_{11}} + {c_{12}}} \right){\varepsilon _c} - 6{B_1}{c_{12}}{c_{44}}{\varepsilon _a} + {A_0}} \right)} 
    \end{aligned}
\end{eqnarray}

\noindent\textbf{(c)} When $z\parallel[001]_{t}$ in the STO substrate ($z\parallel[001]_{c}$), we have $\varepsilon_{xx}=\varepsilon_{yy}=\varepsilon_{c}$, and,
\begin{eqnarray}
    \label{eq:s26c}
    \begin{aligned}
       \omega _{\rm{m}}^{\left[ {001} \right]} = \frac{\gamma }{{{\mu _0}{M_s}}}\sqrt {\frac{{{\mu _0}{M_s}H_x^{{\rm{bias}}} + 2{K_1}}}{{3{c_{11}}{c_{44}}}}} \sqrt { - 6{B_1}{c_{44}}\left( {{c_{11}} + 2{c_{12}}} \right){\varepsilon _c} + {A_0}} 
    \end{aligned}
\end{eqnarray}
\end{subequations}

Since the values of the $\lambda_{100}$ (needed for calculating $B_1$), $\varepsilon_a$ and $\varepsilon_c$ at 80 K remain unknown, we extracted 10 data points from each one of the three experimental FMR curves (see Fig. 4a) and defined the same loss function as that in 250 K, i.e., $\sum {\left( {{\omega _{{\rm{m}},{\rm{cal}}}} - {\omega _{{\rm{m}},{\rm{exp}}}}} \right)^2}$. We applied the limited-memory BFGS algorithm to simultaneously optimize the parameters $\lambda_{100}$, $\varepsilon_a$ and $\varepsilon_c$ via the Python’s SciPy package based on these total 30 data points. Starting from an initial guess of $\lambda _{100}^{{\rm{initial}}} = {10^{ - 5}}$, $\varepsilon _a^{{\rm{initial}}} = \varepsilon _c^{{\rm{initial}}} = 1\% $, the algorithm iteratively minimizes the loss function until convergence, ultimately returning the optimized values of $\lambda_{100}=4.948\times10^{-5}$,  $\varepsilon_{a}=1.008\%$, and $\varepsilon_{c}=0.976\%$. Under these values, based on Eqs. (\ref{eq:s26a})-(\ref{eq:s26c}), one has $\omega _{\rm{m}}^{\left[ {100} \right]} >\omega _{\rm{m}}^{\left[ {001} \right]} >\omega _{\rm{m}}^{\left[ {010} \right]}$.



\newpage

\section{\textbf{S10. Parameters for analytical calculation at 250 K and 80 K}}

\begin{table*}[ht!]
\textbf{\caption{\small Parameters of Analytical Calculation at 250 K ($T > T_\mathrm{S}$)}}
\vspace{5pt}
\centering
\begin{tabular}{l l l}

\hline
LSMO  & Value & Reference\\ 
\hline
$\text{A}_{\rm{ex}}$ (J/m) & $1.41\times10^{-12}$ & Linear interpolation from Ref.~\cite{PhysRevApplied.18.054084} \\
$\text{M}_{\rm{s}}$ (A/m) & $2.283\times10^{5}$ & Experiment measurement \\
K$_1$ (J/m$^3$) & -1766 & Fitted Value\\
$\gamma$ (Hz$\cdot$m/A) & $2.132\times10^{5}$ & Experiment measurement\\
$\lambda_{100}$ &  $3\times10^{-5}$ & Value at 10 K~\cite{ZIESE2002327} \\
$\lambda_{111}$ &  $3\times10^{-5}$ & Same as Value $\lambda_{100}$ at 10 K~\cite{ZIESE2002327} \\
$c_{11}$ (GPa) &  181 & \cite{PhysRevB.57.5093} \\
$c_{12}$ (GPa) &  99 & \cite{PhysRevB.57.5093} \\
$c_{44}$ (GPa) &  56.8 & \cite{PhysRevB.57.5093} \\
B$_1$ (MJ/m$^3$) &  -3.69 & $\frac{3}{2}\lambda_{100}=-\frac{B_1}{c_{11}-c_{12}}$ \\
B$_2$ (MJ/m$^3$) &  -5.11 & $\frac{3}{2}\lambda_{111}=-\frac{B_2}{2c_{44}}$ \\

$a_\mathrm{LSMO}=b_\mathrm{LSMO}=c_\mathrm{LSMO} (\mathring{\mathrm{A}})$ & 3.873& Value at room temperature \cite{doi:10.1126/sciadv.1600245,Nat.Comm.10.1038} \\

\hline
STO  & Value & Reference\\ 
\hline
$c_{44}$ (GPa) &  123 & \cite{PhysRev.129.90} \\
$a_\mathrm{STO}=b_\mathrm{STO}=c_\mathrm{STO}$ $(\mathring{\mathrm{A}})$ & 3.905& \cite{doi:10.1126/sciadv.1600245,Nat.Comm.10.1038} \\

\hline
\end{tabular}
\label{table:s1}

\end{table*}

\vspace{-5pt}
\begin{table*}[ht!]
\textbf{\caption[width=0.8\columnwidth]{\small Parameters of Analytical Calculation at 80 K ($T < T_\mathrm{S}$)}}
\vspace{5pt}
\centering

\begin{tabular}{l l l}
\hline
LSMO  & Fitting & Reference\\ 
\hline
$\text{A}_{\rm{ex}}$ (J/m) & $2.1\times10^{-12}$ & Linear interpolation from~\cite{PhysRevApplied.18.054084} \\
$\text{M}_{\rm{s}}$ (A/m) & $3.727\times10^{5}$ & Experiment measurement \\
K$_1$ (J/m$^3$) & -3647.4 & Linear interpolation from~\cite{ZIESE2002327} \\
$\gamma$ (Hz$\cdot$m/A) & $2.132\times10^{5}$ & Experiment measurement\\
$\lambda_{100}$ &  $4.948\times10^{-5}$ & Fitted Value \\
$\lambda_{111}$ &  $3\times10^{-5}$ & Value at 10 K~\cite{ZIESE2002327} \\
$c_{11}$ (GPa) &  206 & \cite{PhysRevB.57.5093} \\
$c_{12}$ (GPa) &  111 & \cite{PhysRevB.57.5093} \\
$c_{44}$ (GPa) &  60.2 & \cite{PhysRevB.57.5093} \\
B$_1$ (MJ/m$^3$) &  -7.05 & $\frac{3}{2}\lambda_{100}=-\frac{B_1}{c_{11}-c_{12}}$ \\
B$_2$ (MJ/m$^3$) &  -5.42 & $\frac{3}{2}\lambda_{111}=-\frac{B_2}{2c_{44}}$ \\

$a_\mathrm{LSMO}=b_\mathrm{LSMO}=c_\mathrm{LSMO}$ $(\mathring{\mathrm{A}})$ & 3.873& Value at room temperature \cite{doi:10.1126/sciadv.1600245,Nat.Comm.10.1038} \\

\hline
STO  & Value & Reference\\ 
\hline
$c_{44}$ (GPa) &  113 & \cite{PhysRev.129.90} \\
$c_{66}$ (GPa) &  113 & Kept the same as $c_{44}$\\
$a_\mathrm{STO}=b_\mathrm{STO}$ $(\mathring{\mathrm{A}})$ & 3.881& Numerical fitting \\
$c_\mathrm{STO}$ $(\mathring{\mathrm{A}})$ & 3.924& Numerical fitting \\
\hline
\label{table:s2}
\end{tabular}
\end{table*}

\clearpage

\newpage
\section{\textbf{S11. Quantify the magnon-phonon coupling strength at 250 K}}
At 250K, we have ${\varepsilon _{xx}} = {\varepsilon _{yy}} = {\varepsilon ^{mis}} = \frac{{{a_{STO}} - {a_{LSMO}}}}{{{a_{LSMO}}}}\;$. Ignoring the Gilbert damping coefficient $\alpha$ and elastic damping coefficient $\beta$, Eq.(\ref{eq:s15a})-(\ref{eq:s15c}) reduces to:

\begin{subequations}

\begin{eqnarray}
    \label{eq:s27a}
    \begin{aligned}
      {A_{23}} =  - \gamma H_x^{{\rm{bias}}} - \gamma {M_s} - \frac{{2\gamma {K_1}}}{{{\mu _0}{M_s}}} - \frac{{2\gamma B_1^2}}{{3c_{11}^m{\mu _0}{M_s}}} + \frac{{2\gamma B_2^2}}{{c_{44}^m{\mu _0}{M_s}}} + \frac{{2\gamma {B_1}{\varepsilon ^{{\rm{mis}}}}}}{{{\mu _0}{M_s}}} + \frac{{4\gamma {B_1}c_{12}^m{\varepsilon ^{{\rm{mis}}}}}}{{c_{11}^m{\mu _0}{M_s}}}
    \end{aligned}
\end{eqnarray}

\begin{eqnarray}
    \label{eq:s27b}
    \begin{aligned}
      {A_{32}} = \gamma H_x^{{\rm{bias}}} + \frac{{2\gamma {K_1}}}{{{\mu _0}{M_s}}}
    \end{aligned}
\end{eqnarray}

\begin{eqnarray}
    \label{eq:s27c}
    \begin{aligned}
      \;{\vartheta _{yz}} = \frac{{ - \frac{{{B_2}}}{{\omega d}}\left( {4{v^m}{v^{nm}}\;cos\left( {\frac{{L\omega }}{{{v^{nm}}}}} \right){{\sin }^2}\left( {\frac{{d\omega }}{{2{v^m}}}} \right) + {v^{nm}}^2\;sin\left( {\frac{{L\omega }}{{{v^{nm}}}}} \right)\sin \left( {\frac{{d\omega }}{{{v^m}}}} \right)} \right)}}{{c_{44}^m{v^{nm}}\;cos\left( {\frac{{L\omega }}{{{v^{nm}}}}} \right)\sin \left( {\frac{{d\omega }}{{{v^m}}}} \right) + c_{44}^{nm}{v^m}\;cos\left( {\frac{{d\omega }}{{{v^m}}}} \right)\sin \left( {\frac{{L\omega }}{{{v^{nm}}}}} \right)}}
    \end{aligned}
\end{eqnarray}
\end{subequations}

\noindent where ${v^{nm}} = \sqrt {c_{44}^{nm}/{\rho ^{nm}}}$ and ${v^m} = \sqrt {c_{44}^m/{\rho ^m}}$ are the transverse sound velocities in the non-magnetic (NM) and magnetic (M) layers, respectively. Note that the denominator of $\Omega_{y} (z,t)$ equals zero is the resonance condition for the standing acoustic phonon in M/NM bilayer under stress-free boundary conditions described above. In the limit of $L\gg d$, the acoustic phonon in the M/NM bilayer can be approximated as the acoustic phonon in a single NM layer with thickness L, with eigenmodes $\omega _n^0 = n\pi {v^{nm}}/L$. Thus, after expanding near $\omega  \approx \omega _n^0$, Eq.(\ref{eq:s27c}) further reduces to:

\begin{eqnarray}
    \label{eq:s28}
    \begin{aligned}
      {\vartheta _{yz}} \approx \frac{{ - {B_2}L{\rho ^{nm}}\left( {\omega  - \omega _n^0} \right)}}{{c_{44}^md{\rho ^m}\left( {\omega  - \omega _n^0} \right)}}
    \end{aligned}
\end{eqnarray}

where both denominator and numerator are small quantities. Note that the susceptibility tensor $\boldsymbol{\chi}$ in Eq.(\ref{eq:s16b}) can be reformatted as:

\begin{eqnarray}
    \label{eq:s29}
    \begin{aligned}
        {\bf{\chi }} = {\left( {\begin{array}{*{20}{c}}
        {\frac{{i\omega }}{\gamma }}&0&0\\
        0&{\frac{{{A_{32}}}}{\gamma }}&{\frac{{i\omega }}{\gamma }}\\
        0&{ - \frac{{i\omega }}{\gamma }}&{\frac{{{B_2}{\vartheta _{yz}}}}{{{\mu _0}{M_s}}} - \frac{{{A_{23}}}}{\gamma }}
        \end{array}} \right)^{ - 1}}
    \end{aligned}
\end{eqnarray}
At the strongest absorption, the susceptibility reaches its local maximum. Thus, we take $\det\left({\boldsymbol{\chi}^{-1}}\right) = 0$ which eventually reduces to a quadratic equation of $\omega$:

\begin{eqnarray}
    \label{eq:s30}
    \begin{aligned}
        A{\omega ^2} + B\omega  + C = 0
    \end{aligned}
\end{eqnarray}

\noindent where $A = c_{44}^md{\mu _0}{M_s}{\rho ^m}\omega _n^0$, $B = \left( {{A_{23}}{A_{32}} - \omega {{_n^0}^2}} \right)c_{44}^md{\mu _0}{M_s}{\rho ^m} + {A_{32}}B_2^2L\gamma {\rho ^{nm}}$, $C =  - {A_{23}}{A_{32}}\\c_{44}^m d {\mu _0}{M_s}{\rho ^m}\omega _n^0 - {A_{32}}B_2^2L\gamma {\rho ^{nm}}\omega _n^0$.

Therefore, the magnon-phonon coupling strength $\Omega$ is expressed as the half-width of the frequency gap near the center frequency representing the n-th TA mode ($\omega  \approx \omega _n^0)$:

\begin{eqnarray}
    \label{eq:s31}
    \begin{aligned}
        {\rm{\Omega }} = \frac{{\sqrt {{B^2} - 4AC} }}{{2A}} = \frac{{\left| {{A_{23}}{A_{32}}c_{44}^md{\mu _0}{M_s}{\rho ^m} + {A_{32}}B_2^2L\gamma {\rho ^{nm}} + c_{44}^md{\mu _0}{M_s}{\rho ^m}\omega {{_n^0}^2}} \right|}}{{2c_{44}^md{\mu _0}{M_s}{\rho ^m}\omega _n^0}}
    \end{aligned}
\end{eqnarray}

Ignoring the anisotropy coefficient $K_1$ and the normal strain magnetoelastic coupling coefficients $B_1$, the above equation further reduces to:

\begin{eqnarray}
    \label{eq:s32}
    \begin{aligned}
        {\rm{\Omega }} = \frac{{B_2^2L{\rho ^{nm}}{\gamma ^2}\left( {B_2^2 - c_{44}^m{\mu _0}M_s^2 + \sqrt {{{\left( {B_2^2 - c_{44}^m{\mu _0}M_s^2} \right)}^2} + {{\left( {\frac{{2c_{44}^m{\mu _0}{M_s}\omega _n^0}}{\gamma }} \right)}^2}} } \right)}}{{4c{{_{44}^m}^2}d\mu _0^2M_s^2{\rho ^m}\omega _n^0}}
    \end{aligned}
\end{eqnarray}

Note that setting the shear strain magnetoelastic coupling coefficients $B_2$ also to zero leads to $\Omega=0$, indicating no coupling between magnon and phonon.

Compared with the expression in Dr. O. Klein's paper \cite{PhysRevB.101.060407}:

\begin{eqnarray}
    \label{eq:s33}
    \begin{aligned}
        {{\rm{\Omega }}^{{\rm{Klein}}}} = \frac{B}{{\sqrt 2 }}\sqrt {\frac{\gamma }{{{\omega _s}{M_1}sd}}} \left( {1 - \cos \left( {{\omega _s}\frac{d}{v}} \right)} \right)
    \end{aligned}
\end{eqnarray}

our expression in Eq.(\ref{eq:s31}) only deals with the case when the thickness of NM layer is far larger than that of the M layer ($L\gg d$) but shows the dependency of the coupling strength with respect to much more material parameters, e.g., anisotropy coefficient $K_1$, the elastic stiffness coefficients $c_{ij}$, the mass densities of both M ($\rho^m$) and NM ($\rho^{nm}$) layers, the lattice mismatch strain between the M and NM layers $\epsilon^{\rm{mis}}$, etc. 

Besides, the expression in Eq.(\ref{eq:s31}) includes both normal ($B_1$) and shear ($B_2$) strain magnetoelastic coupling coefficients and deals with them more properly when they have negative values. The FMR relation between the magnon frequency and the external bias field is also treated properly with all the effective field terms included, instead of just using the linear approximation given by the Kittel equation $\omega  \approx \gamma {\mu _0}\left( {{\rm{H^{bias}}} - {M_s}} \right)$.

Take $d=38$ nm and plug in the materials parameters of LSMO and STO under 250 K to Eq.(\ref{eq:s31}), we obtain the following relation between the coupling strength and the standing phonon frequency, where the red stars indicate experiment measurements (see Fig. 4(d) in the main text).

\newpage

\vspace{-5pt}

\clearpage


\clearpage

\bibliographystyle{apsrev4-1}
\nocite{apsrev41Control}

\bibliography{Supplemental.bib}